\documentclass[%
 preprint,
nofootinbib,
 amsmath,amssymb,
 aps,
]{revtex4-1}

\usepackage{comment}
\usepackage{graphicx}
\usepackage{grffile}
\usepackage{epstopdf}
\usepackage{enumerate}
\usepackage{tensor}
\usepackage{braket}       
\usepackage{bm}
\usepackage{slashed}
\usepackage{appendix}

\newcommand{\tr}{\text{tr }}

\begin{document}
\title{Spectrum of a Gross-Neveu Yukawa model with flavor disorder in three dimensions}
\author{Shiroman Prakash}

\affiliation{
Dayalbagh Educational Institute,  Agra- 282005, India}

\date{\today}

\begin{abstract}  We show that a variant of the Gross-Neveu Yukawa model with disorder provides a real, nonsupersymmetric generalization of the Sachdev-Ye Kitaev (SYK) model to three dimensions. The model contains $M$ real scalar fields and $N$ Dirac (or Majorana) fermions, interacting via a Yukawa interaction with a local Gaussian random coupling in three dimensions. In the limit where $M$ and $N$ are both large, and the ratio $M/N$ is held fixed, the model defines a line of infrared fixed points parametrized by $M/N$, reducing to the  Gross-Neveu vector model when $M/N=0$. When $M/N$ is nonzero, the model is dominated by melonic diagrams and gives rise to SYK-like physics. We compute the spectrum of single-trace operators in the theory, and find that it is real for all values of $M/N$. 
\end{abstract}

\maketitle 

\section{Introduction}

The large-$N$ limit \cite{'tHooft:1973jz} is a powerful tool to understand strongly interacting quantum field theories. In this limit, unexpected, classical descriptions can emerge --  the most notable example of which is classical gravity in anti-de Sitter space \cite{Maldacena:1997re, Gubser:1998bc, Witten:1998qj}. 

Originally, two large $N$ limits were known  -- the large $N$ limit of theories with dynamical fields in adjoint/matrix representations, exemplified by \cite{'tHooft:1973jz}, in which all planar Feynman diagrams contribute; and theories with dynamical fields in the vector representation, exemplified by \cite{'tHooft:1974hx}, in which a summable subset of planar Feynman diagrams contributes. Large $N$ vector models typically possess a slightly broken higher-spin symmetry, and any dual gravitational description for such a conformal field theory (CFT) would include a tower of massless higher-spin gauge fields \cite{KlebanovPolyakov, Sezgin:2003pt,  Leigh:2003gk, Giombi:2011kc, Aharony:2011jz, Maldacena:2011jn, MZ}. On the other hand, strongly-interacting large $N$ matrix models are, at least in some cases with sufficient supersymmetry \cite{Maldacena:1997re, ABJM}, believed to be dual to theories of Einstein supergravity. 

Theories also exist whose large $N$ limit interpolates between a vector model and a matrix model. The ABJ theory \cite{ABJ:2008gk} is a bifundamental $U(N)\times U(M)$ Chern-Simons theory that can be studied in the limit where $M$ and $N$ are both large and the ratio $M/N$ is fixed. When $M/N \ll 1$, the theory possesses a slightly-broken higher-spin symmetry and is effectively a vector model, dual to a higher-spin gauge theory (with $U(M)$ Chan-Paton indices) \cite{ABJTriality}. When $M/N=1$, the theory becomes ABJM theory \cite{ABJM} and is dual to type IIA supergravity in $AdS_4 \times CP_3$ at strong coupling. The parameter $M/N$ can be understood as a gravitational 't Hooft coupling in the bulk; as this parameter is increased from zero to one, the higher spin gauge fields somehow coalesce into strings.

Large $N$ vector model CFTs without supersymmetry are common. Do there exist large $N$ CFTs without supersymmetry dual to Einstein gravity in AdS? A variant of the weak gravity conjecture \cite{Ooguri:2016pdq} suggests that the answer is no. However, motivated by ABJ \cite{ABJTriality}, one may promote nonsupersymmetric vector models to bifundamental theories, and study their behavior as a function of $M/N$  \cite{Gurucharan:2014cva, Kapoor:2021lrr}. However, all examples studied so far become complex at some critical  value of $M/N$, as can be seen via the epsilon expansion \cite{Kapoor:2021lrr, Osborn:2017ucf}, or, in the case of certain matter Chern-Simons theories \cite{Gurucharan:2014cva, GuruCharan:2017ftx}, cannot be studied at strong coupling. 

Recently, a new large $N$ limit -- the melonic limit\footnote{This limit was first understood via tensor models, e.g., \cite{Gurau:2009tw, Gurau:2011aq, Gurau:2011xq, Bonzom:2011zz, Tanasa:2011ur, Bonzom:2012hw, Carrozza:2015adg, Witten:2016iux, Klebanov:2016xxf, Prakash:2019zia}} -- has emerged, via the study of the Sachdev-Ye Kitaev (SYK) model \cite{Sachdev:1992fk,KitaevTalk,Maldacena:2016hyu, Kitaev:2017awl, Rosenhaus:2019mfr}, a chaotic model in one dimension, exhibiting features of black hole physics, see, e.g., \cite{Fu:2016vas, Jevicki:2016bwu, Mandal:2017thl, Das:2017pif, Das:2017hrt, Bulycheva:2017ilt, Gross:2017hcz}. The melonic limit is dominated by a  summable subset of planar diagrams that  manages to capture non-trivial aspects of gravitational physics. 
Does there exist a nonsupersymmetric higher-dimensional generalization of the SYK model \cite{Turiaci:2017zwd, Murugan:2017eto, Liu:2018jhs, Peng:2018zap, Chang:2021fmd, Chang:2021wbx}? All nonsupersymmetric constructions, \cite{Murugan:2017eto, Liu:2018jhs, Chang:2021wbx} encounter an operator with a complex scaling dimension. Similar results have also been observed in tensor models, \cite{Klebanov:2016xxf, Giombi:2017dtl, Prakash:2017hwq, GKPPT}. In particular, \cite{ Chang:2021wbx}, inspired by \cite{Peng:2018zap}, defined a ``biconical'' SYK model that interpolates between the critical $O(N)$ vector model when $M/N \ll 1$ and a model with SYK like physics when $M/N$ is finite. Their model becomes complex when $M/N$ exceeds $0.22$, reminiscent of bifundamental critical scalars in $d=3$\cite{Kapoor:2021lrr, Osborn:2017ucf}. By contrast,  \cite{Chang:2021wbx} shows that a supersymmetric version of their theory is real for all $M/N$. 

It is natural to expect that nonsupersymmetric melonic CFTs do not exist -- and, if this were true, it would provide strong support for the conjecture that nonsupersymmetric strongly interacting CFTs dominated by planar diagrams do not exist. Here, we show that this expectation is false by computing the spectrum a simple example of a nonsupersymmetric CFT, recently proposed in \cite{Kim:2020jpz}, dominated by melonic diagrams that can be studied for all values of $M/N$. 

\section{Gross-Neveu Yukawa model with disorder}

A starting point in a search for higher-dimensional SYK physics is the Gross-Neveu (GN) model \cite{Gross:1974jv}, which possesses a non-trivial ultraviolet fixed point in $d=2+\epsilon$, that, at least when $N$ is large, defines a conformal field theory in $d=3$. A tensorial GN model was constructed in \cite{Prakash:2017hwq, Benedetti:2017fmp,Benedetti:2019eyl}, but its spectrum of operators includes one with a complex dimension -- an identical spectrum arises in a GN model with a random 4-fermion interaction. However, there exists a cousin of the GN model -- the Gross-Neveu Yukawa (GNY) model \cite{Hasenfratz1991TheEO, ZINNJUSTIN1991105, Moshe:2003xn, Fei:2016sgs} -- that gives rise to an \textit{infrared} fixed point in $d=4-\epsilon$, equivalent to the GN fixed point when $N$ is large \cite{Moshe:2003xn, Fei:2016sgs}. See, e.g., \cite{Muta:1976js, Wetzel:1984nw, Gracey:1990wi, Zinn-Justin:1991ksq, Gracey:1991vy, Luperini:1991sv, Vasiliev:1992wr, Gracey:1992cp,  Kivel:1993wq, Gracey:1993kc,  Derkachov:1993uw, Gracey:2008mf, Raju:2015fza, Ghosh:2015opa, Manashov:2016uam, Gracey:2016mio, Giombi:2017rhm, Cresswell-Hogg:2022lgg} for more on the GN and GNY theories. 

Inspired by \cite{Chang:2021wbx}, we study a variant of the GNY model constructed by \cite{Kim:2020jpz}. The model, which is closely related to several one-dimensional models generalizing SYK  \cite{Kang:2021gsx, Kim:2019lwh, Wang:2019bpd, Esterlis:2019ola, Marcus:2018tsr, Bi:2017yvx},  consists of $M$ real scalar fields, $\sigma^a$, $a=1,~\ldots, M$, and $N$ Dirac or Majorana fermions, $\psi^i$, $i=1,~\ldots, N$ with the interaction
, 
\begin{equation}
    \mathcal S_{int} = \int d^d x ~g_{ai}{}^{j}\sigma^a \bar{\psi}^i \psi_j,
\end{equation}
which is relevant for $d<4$, and leads to an IR fixed point in $d=3$.  


$g_{ai}{}^{j}$ is a random coupling with quenched disorder, with zero mean $ \langle g_{ai}{}^{j} \rangle =0$, but non-zero variance
\begin{equation}
    \langle g_{ai}{}^{j}g_{bk}{}^{l} \rangle = \frac{J}{N^2} \delta_{ab} \delta_{i}^l \delta_{k}^j.
\end{equation}
The resulting average is invariant under $O(M)\times U(N)$ for the theory with Dirac fermions (or $O(M)\times O(N)$ for  Majorana fermions.) 

We take both $M$ and $N$ to be large, keeping the ratio $\lambda=\frac{M}{N}$ fixed. It follows from \cite{Bonzom:2018jfo} that this limit is dominated by melonic diagrams.

When $\lambda=0$, the theory reduces to the GN model, but it exhibits SYK-like physics for $M/N$ finite. 

Does the spectrum of single-trace primaries, like all other known nonsupersymmetric, higher-dimensional melonic models, include a complex scaling dimension for $M/N$ sufficiently large?
We compute the spectrum  in $d=3$, and find it is real for all values of $M/N$.

We present the spectrum for the theory with $N$ Dirac fermions below -- the spectrum of the theory with $2N$ Majorana fermions is a truncation to operators with even spin.

\section{Exact propagators}
The exact propagators for $\psi$ and $\sigma$ were obtained in \cite{Kim:2020jpz}, as we review below.  
\begin{figure}
    \centering
    \includegraphics[width=0.7\textwidth]{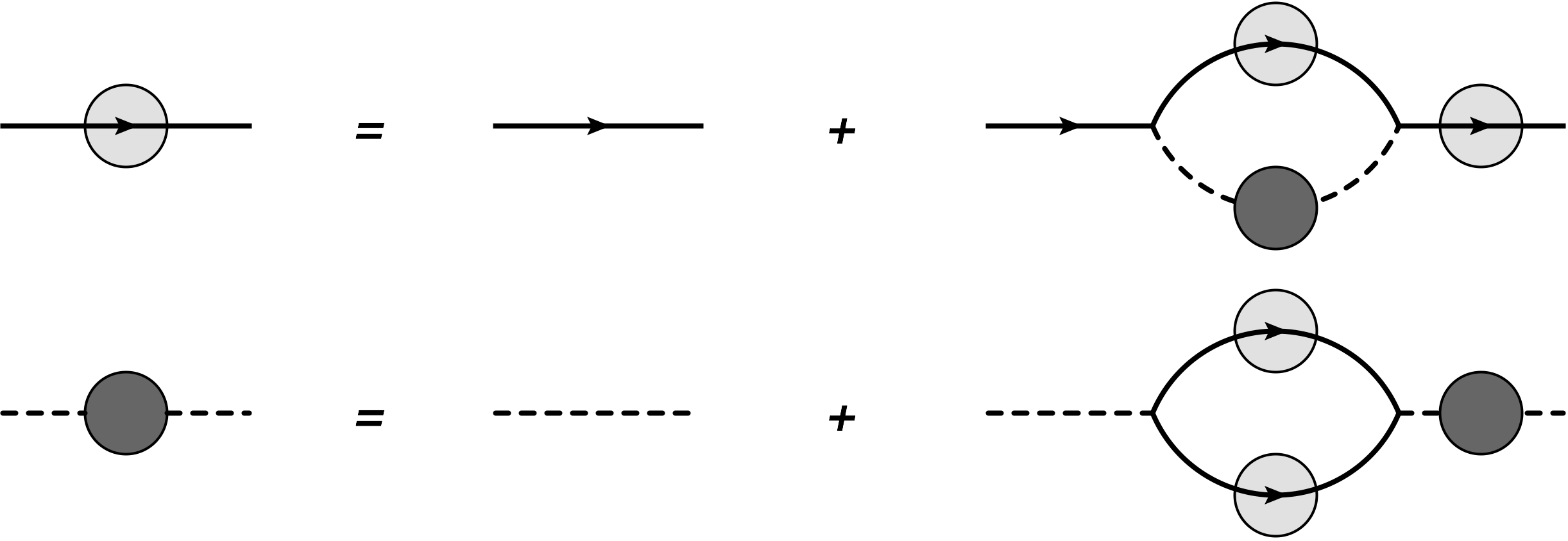}
    \caption{The Schwinger-Dyson equations for the exact propagators. Solid and dashed lines denote fermion and scalar propagators respectively.}
    \label{fig:gap-eqn}
\end{figure}

Denote the exact propagators  by,
\begin{equation}
    \langle \psi^i(p)\bar{\psi}_j(-q) \rangle = (2\pi)^d G(p) \delta^d(p-q) \delta^i_j, ~ \langle \sigma^a(p) \sigma^b(-q) \rangle = (2\pi)^d G(p) \delta^d(p-q) \delta^{ab}.
\end{equation}
In the melonic limit, these propagators satisfy the Schwinger-Dyson equations illustrated in Figure \ref{fig:gap-eqn},  
\begin{equation}
    F_0(p)^{-1}=F(p)^{-1} - J\int \frac{d^dq}{(2\pi)^d} \tr \left( G(q)G(p+q) \right) \label{gapf}
\end{equation}
\begin{equation}
    G_0(p)^{-1} = G(p)^{-1} + J\lambda \left(\int \frac{d^dq}{(2\pi)^d} G(p-q) F(q)\right) \label{gapg}.
\end{equation}
where $G_0(p)=\frac{1}{i \slashed{p}}$ and $F_0(p)=\frac{1}{p^2}$ .

The conformal ansatz for interacting propagators in the IR is 
\begin{equation}
    G(p)=A \frac{i \slashed{p}}{p^{d-2\Delta_\psi+1}}, ~ F(p)= \frac{B}{p^{d-2\Delta_\sigma}}. \label{conformal-ansatz}
\end{equation}

For an interacting IR solution, we require, \begin{equation}2\Delta_\psi+\Delta_\sigma=d,\end{equation} and \begin{equation}
(d+2)/4>\Delta_\psi>(d-1)/2.
\end{equation} 

\begin{figure}
    \centering
    \includegraphics[width=0.7\textwidth]{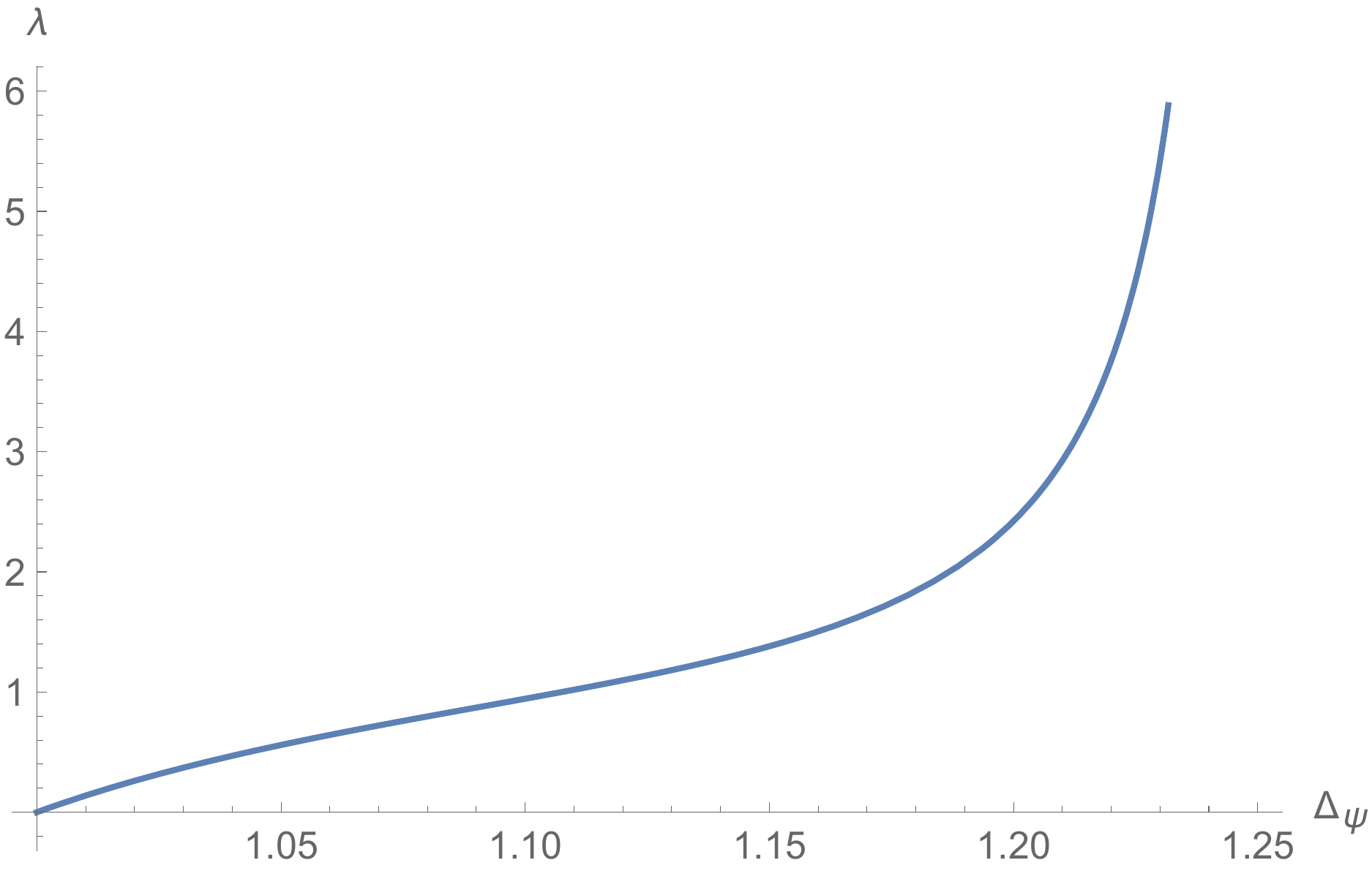}
    \caption{The gap equation for $\Delta_\psi$ in $d=3$ possesses a unique solution for all $\lambda$.}
    \label{fig:gap-GNY}
\end{figure}

Evaluating equations \eqref{gapf} and \eqref{gapg} using \eqref{conformal-ansatz}, we obtain
an equation that determines $\Delta_\psi$, which, in $d=3$, reads,
\begin{equation}
    \lambda=\frac{8 (2 \Delta_\psi -5) (2 \Delta_\psi -3) \sin ^3(\pi  \Delta_\psi ) \cos (\pi  \Delta_\psi ) \Gamma (3-4 \Delta_\psi ) \Gamma (4 \Delta_\psi -4)}{\pi }.
    \label{gap-eqn-3d}
\end{equation} Figure \ref{fig:gap-GNY} shows that, in $d=3$, \eqref{gap-eqn-3d} has a unique solution within the allowed range, for all $\lambda$. 
For small $\lambda$,
\begin{equation}
    \Delta_\psi = 1+\frac{2 \lambda }{3 \pi ^2}+\frac{80 \lambda ^2}{27 \pi ^4} - \frac{\left(120 \pi ^2-4384\right) \lambda ^3}{243 \pi ^6}+O(\lambda^4).
\end{equation}
This agrees with the $O(1/N)$ correction to $\Delta \psi$ in the GN model.

As $\lambda \to \infty$, $\Delta_\psi \to 5/4$ and $\Delta_\sigma \to 1/2$. In between, there is a ``critical value'' of $\lambda_*=\frac{2 (d-1) {d_\gamma}}{d+2}=\frac{8}{5}$ defined by the condition $2\Delta_\sigma=2\Delta_\psi-1$, for which $\Delta_\sigma=2/3$ and $\Delta_\psi = 7/6$.

Does the free theory flow to this putative IR limit? In Appendix \ref{numerics}
 we solved the full gap equations \eqref{gapf}-\eqref{gapg}, for $\lambda=0$ and $\lambda=\infty$, and found flows from the free theory to the IR solution. Presumably, similar flows also exist for intermediate values of $\lambda$. 

\section{Operator spectrum}

\begin{figure}
    \centering
    \includegraphics[width=0.7\textwidth]{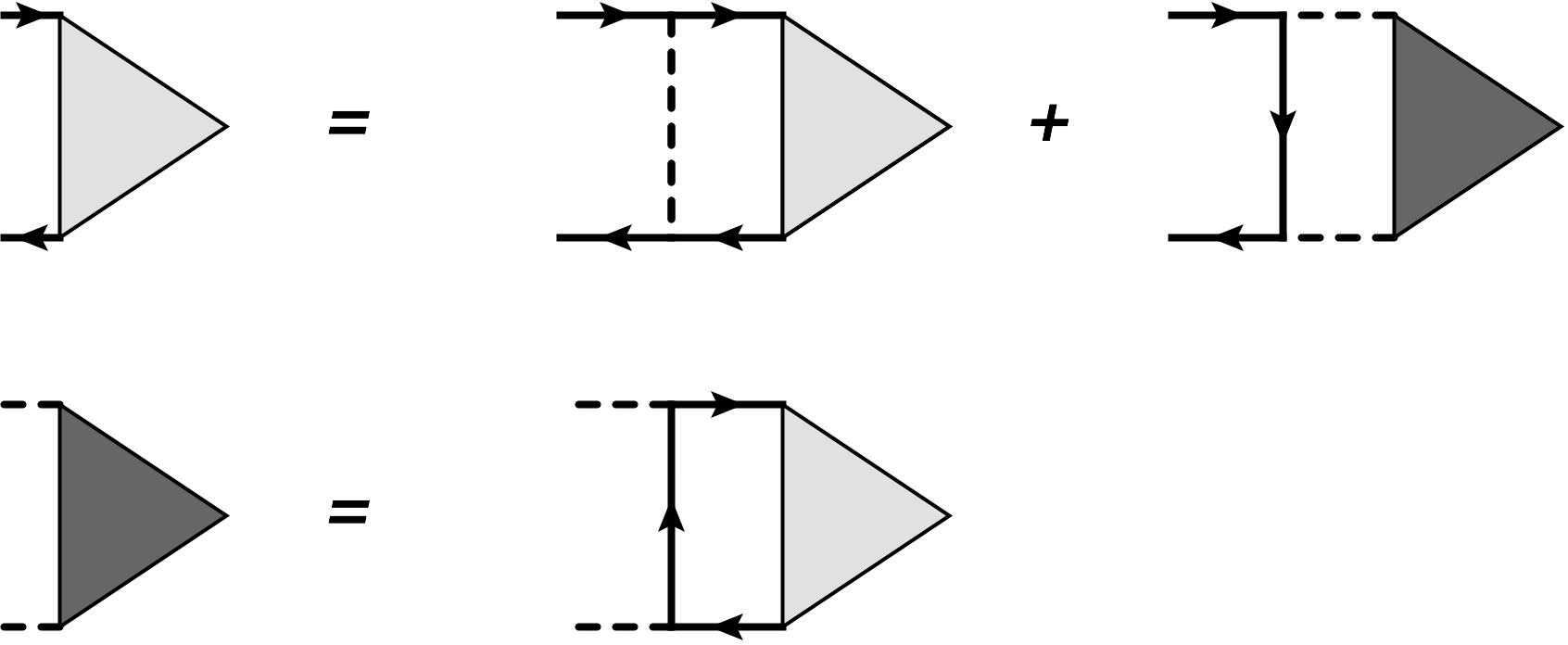}
    \caption{In the melonic IR limit, the exact three-point functions $\langle \sigma(x_1) \sigma(x_2) \mathcal O_{\tau,s}(x_3) \rangle$ and $\langle \psi(x_1) \bar{\psi}(x_2) \mathcal O_{\tau,s} \rangle(x_3)$ (depicted as dark and light gray triangles) satisfy the ladder equation depicted. All propagators are exact.}
    \label{kernel-figure}
\end{figure}
The single-trace primaries are bilinear singlets of the $O(M)\times U(N)$ symmetry group.\footnote{Note that bilinears such as $\sigma_a \psi_i$ or, more generally $\sigma_a A^{ai} \psi_i$, for any $M \times N$ matrix $A$, are not invariant under the $O(M)\times U(N)$ symmetry group -- anomalous dimensions of such operators are suppressed by $1/N$.} In the free theory, these are,
\begin{enumerate}
    \item spin-$s$ bilinears of $\psi$,  of the schematic form, 
$\bar{\psi}^i (\partial \cdot \gamma )^n \psi_i,$ for $s=0$, and, 
$\bar{\psi}^i \gamma \cdot z (\partial \cdot z)^{s-1} (\partial \cdot \gamma )^n \psi_i,$  for $s>0$;
    \item and (even) spin-$s$ bilinears of $\sigma$, of the schematic form $$\sigma^a (\partial \cdot z)^s (\partial^{2m}) \sigma^a.$$ 
\end{enumerate}   Here, $z$ denotes a null polarization vector (see, e.g., \cite{GPY, Giombi:2016hkj}). 
In the IR fixed point, fermion and scalar bilinears which are parity-even can mix.



 Let $\mathcal O_{s,\tau}$ be an operator with well-defined scaling dimension, and let $\tau=\Delta_{\mathcal O} -s$ be its twist. In the IR limit, the three-point functions $\langle \sigma(x_1) \sigma(x_2) \mathcal O_{\tau,s}(x_3) \rangle$ 
and $\langle \psi(x_1) \bar{\psi}(x_2) \mathcal O_{\tau,s} (x_3) \rangle$  satisfy the melonic ladder equation, shown schematically in Figure \ref{kernel-figure}. Following \cite{Polchinski:2016xgd}, we use this equation, along with conformal invariance, to determine the allowed operator-spectrum in the IR. 
Details  are in Appendix \ref{app-integration-kernel}.
.

Figure \ref{kernel-figure} translates into condition that an integration kernel (which is, in general, a matrix, due to mixing and multiple allowed forms for conformally-invariant three-point functions), $\mathbf{K}_s^{\text{even/odd}}(\Delta)$, defined for parity even/odd spin $s$ operators, has eigenvalue $1$. 

\section{Spectrum in $d=3$}
We now show that the spectrum of single-trace operators is real for all $\lambda$ in $d=3$.  As usual for SYK-like models, the results involve some numerical solutions of transcendental equations. Alternatively, analytic expressions for the spectrum can be obtained in $d=4-\epsilon$, which are presented in Appendix \ref{app:epsilon}.

\subsection{Parity-odd scalars}

We expect the scaling dimensions of parity-odd scalars, denoted as $\tilde{\Delta}_n$, to take the form $\tilde{\Delta}_n=2\Delta_\psi+2n+\tilde{\gamma}_{0,n}$, with $\tilde{\gamma}_{0,n} \to 0$ as $n \to \infty$.

Figure \ref{kernel-figure} translates into the equation
\begin{equation}
    K_0^{\text{odd}}(\tilde{\Delta}) = 1, \label{parity-odd-eigenvalue-eqn}
\end{equation}
as described in Appendix \ref{app-integration-kernel}.

\begin{figure}
    \centering
    \includegraphics[width=0.7\textwidth]{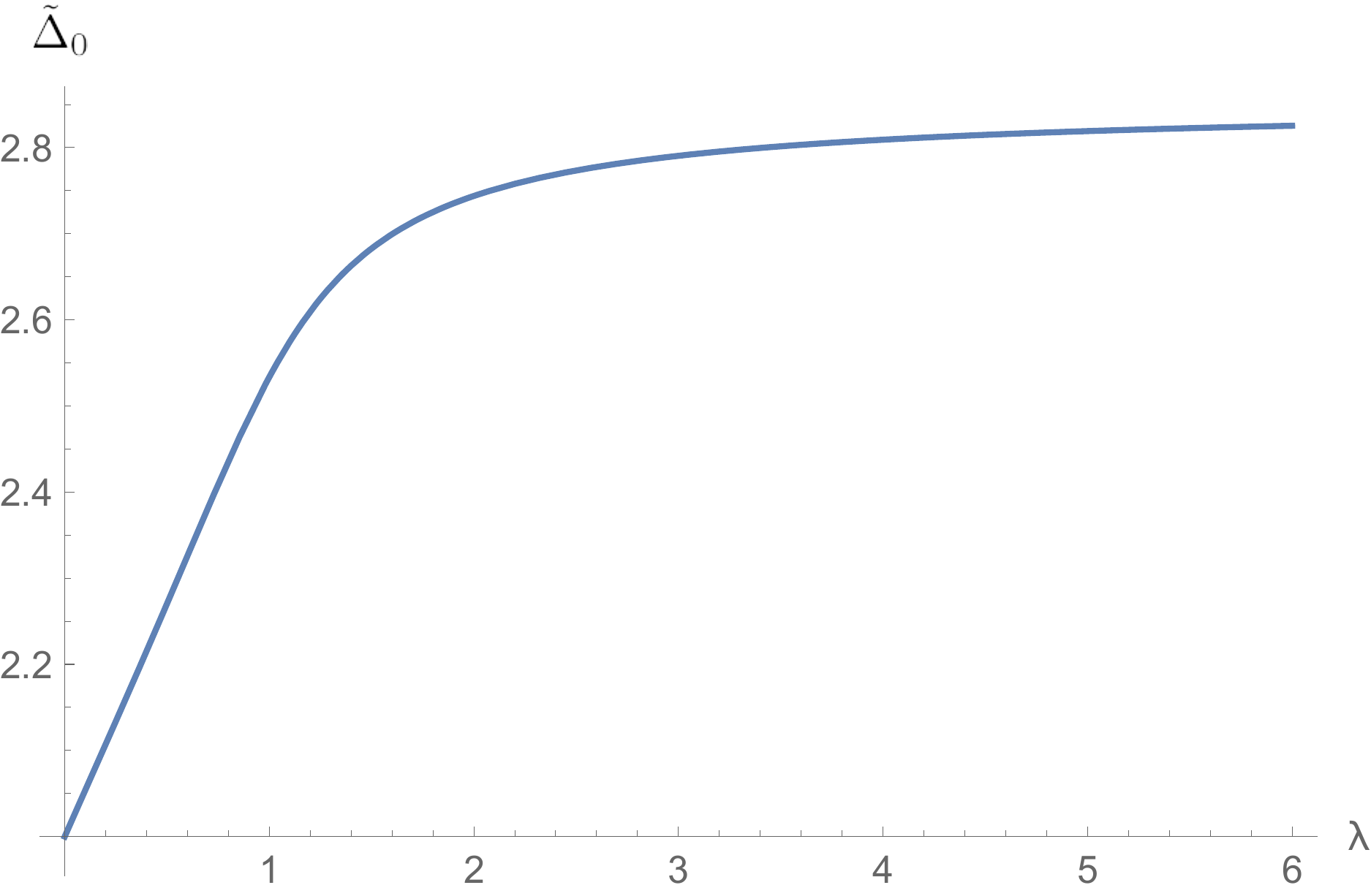}
    \caption{Scaling dimension of the lowest-twist parity-odd scalar, $\bar{\psi}\psi$, as a function of $\lambda$.}
    \label{fig:parity-odd-scalar}
\end{figure}

For $\lambda=1$, we find $\tilde{\Delta}_n =2.53354$, $4.25934$, $6.23292$, $8.22512$,  $\ldots$, approaching $2\Delta_\psi+2n=2.21492+2n$.  
For $\lambda=\infty$, we find $\tilde{\Delta}_n =2.85171$, $4.66395$, $6.60305$, $8.57464$,  $\ldots$, also approaching $2\Delta_\psi+2n=2.5+2n$. 
The behavior for intermediate values of $\lambda$ is similar, and, the spectrum is \textbf{real} for all values of $\lambda$. The scaling dimension of the lowest-dimension parity-odd scalar is shown in Figure \ref{fig:parity-odd-scalar}. In the melonic GN model, \cite{Prakash:2017hwq}, this operator has a complex scaling dimension. 

For small $\lambda$,  
\begin{equation}
    \tilde{\Delta}_0=2+\frac{16}{3 \pi ^2}\lambda -\frac{128}{27 \pi ^4} \lambda^2 + O(\lambda^3) = 2 \Delta_\psi + \frac{4 \lambda }{\pi ^2} + O(\lambda^2),
\end{equation} 
\begin{eqnarray}
\tilde{\Delta}_n & = & 2 \Delta_\psi + 2n + \frac{16}{3 \pi ^4 n (2 n+1)} \lambda^2  + \nonumber \\
&& \left( \frac{192 (\log 2n +\gamma +2)+32}{27 \pi ^6 n^2}+O\left(\frac{1}{n^3}\right) \right) \lambda^3 + O(\lambda^4)  \label{higher-twist-parity-odd-scalar-spectrum}
\end{eqnarray}
At large $n$, 
\begin{equation}
    \tilde{\gamma}_{0,n} \sim \frac{1}{n^{2\Delta_\sigma}}.
\end{equation}

\subsection{Parity-even scalars}
The spectrum of parity-even scalars includes mixtures of scalar and fermion bilinears, operators, denoted as type-A and type-B,  whose scaling dimensions are of the form $\Delta_{n}^A=2\Delta_\psi+2n+1+\gamma^A_{0,n}$ and $\Delta_{n}^B=2\Delta_\sigma+2n+\gamma^B_{0,n}$, respectively, where $\gamma^{A/B}_{0,n} \to 0$ as $n \to \infty$.

Due to mixing, the integration kernel for parity-even scalars is a $2\times 2$ matrix, $\mathbf K_{0}^{\text{even}}(\Delta)$, given in  \eqref{parity-even-scalar-kernel} in Appendix \ref{app-integration-kernel}. 

The condition 
\begin{equation}
    \det \left(\mathbf K_{0}^{\text{even}}(\Delta)-\mathbf 1 \right)=0, \label{parity-even-kernel-equation}
\end{equation} 
determines the allowed values of $\Delta$.

\begin{figure}
    \centering
    \includegraphics[width=0.4\textwidth]{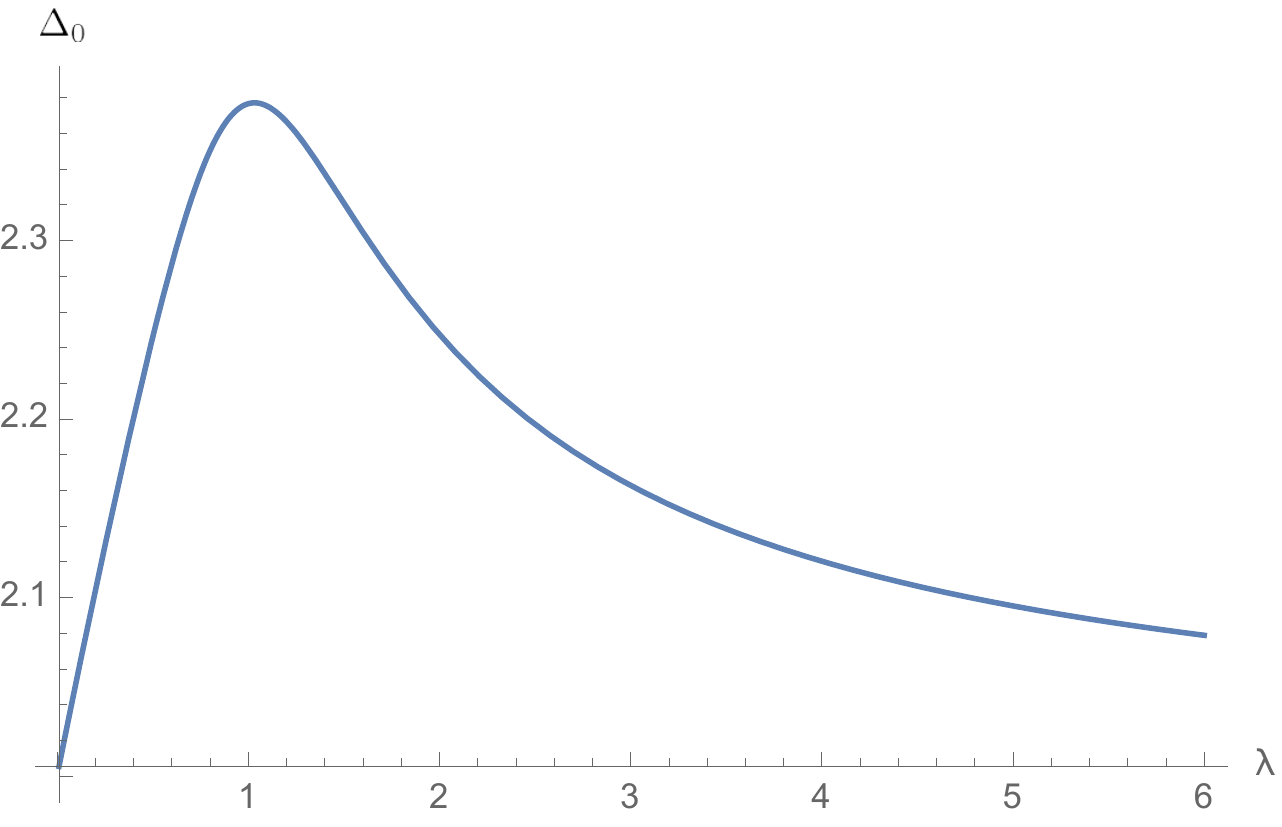}
    \caption{The scaling dimension $\Delta_0(\lambda)$ of the lowest-dimension parity-even scalar as a function of $\lambda$. $\Delta(\lambda)$ attains its maximum, $2.377$ near $\lambda=1.03$.}
    \label{fig:parity-even-scalar}
\end{figure}

We find the spectrum for $\lambda=1$ is $2.37666$, $3$, $3.70347$, $5.23918$, $5.57576$, $7.22808$, $7.57099$, $\ldots$. The scaling dimension of the lowest operator is plotted as a function of $\lambda$ in Figure \ref{fig:parity-even-scalar}.\footnote{Both the lowest scaling dimension $\Delta_0$ and its shadow $\Delta_0'=3-\Delta_0$ are above the unitary bound for all $\lambda$. Our analysis does not determine which is realized in the IR limit; however, as $\lambda \to 0$, ${\Delta_0} \to 2^+=2\Delta_\sigma$, in agreement with the GN model. Assuming the spectrum is a continuous function of $\lambda$, we conclude $\Delta_0$ is the scaling dimension in the IR.} 

\begin{figure}
    \centering
    \includegraphics[width=0.7\textwidth]{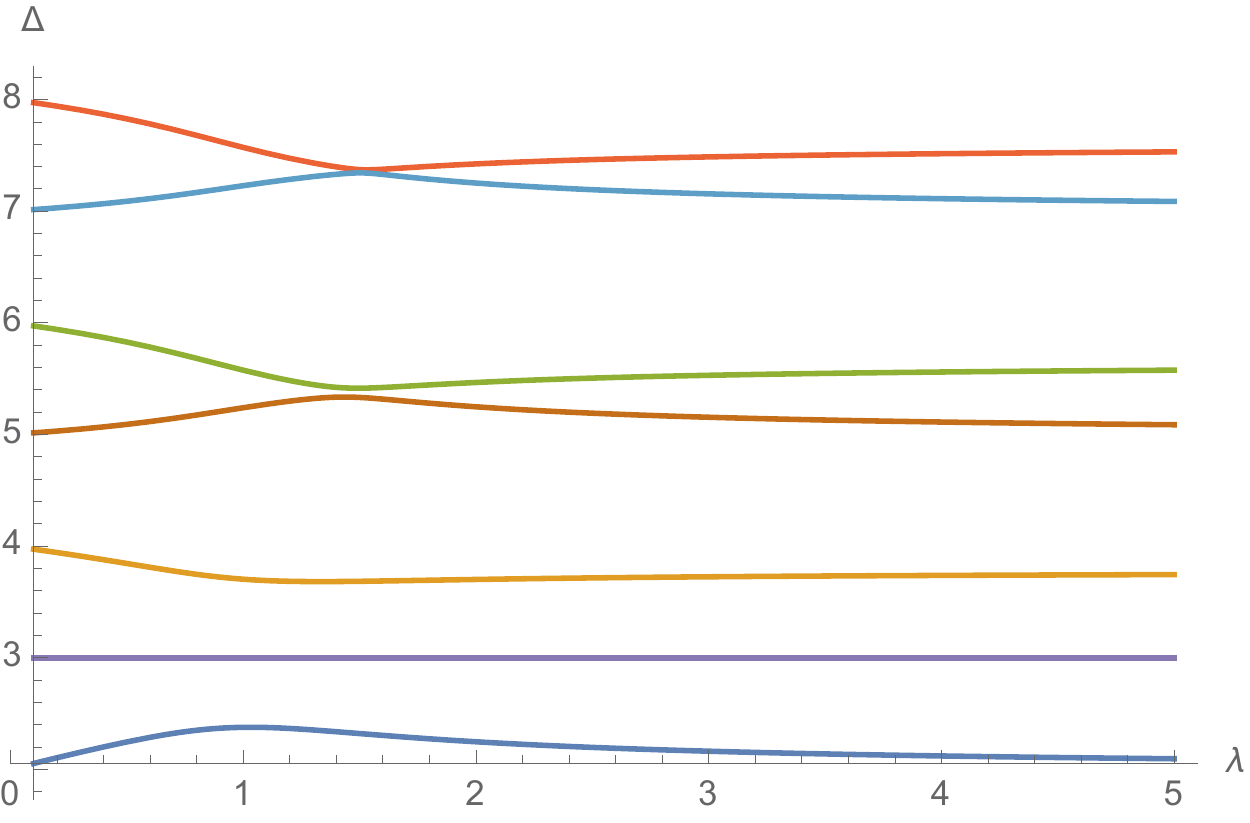}
    \caption{The lowest 8 scaling dimensions of parity-even scalars a function of $\lambda$. $\Delta_0(\lambda)$ attains its maximum, $2.377$ near $\lambda=1.03$. The kink in the graphs is, for large $n$, near  $\lambda=8/5$.}
    \label{fig:higher-n-parity-even-scalars}
\end{figure}

We plot the scaling dimensions of the 8 lowest scalar primaries in Figure \ref{fig:higher-n-parity-even-scalars}. As Figure \ref{fig:higher-n-parity-even-scalars} illustrates, the spectrum is real for all values of $\lambda$. As $n \to \infty$, the scaling dimensions approach $2\Delta_\sigma+2n$ and $2\Delta_\psi+2n+1$.

There exists an exactly marginal operator with $\Delta=3$, for all $\lambda.$ We expect that this corresponds to a mixture of $\sigma^a \partial^2 \sigma^a$ and $\bar{\psi}^i \slashed{\partial} \psi_i$,  which is redundant and vanishes by the equation of motion. as in \cite{Bulycheva:2017ilt, Gross:2016kjj}.


For small $\lambda$, 
\begin{eqnarray}
\Delta^A_0 & = & 3, 
\end{eqnarray}
\begin{equation}
\begin{split}
    \Delta^A_n & =  2\Delta_\psi + 2n + 1 + \frac{16 \left(2 n^2+3 n-1\right)}{3 n \left(4 n^3+12 n^2+11 n+3\right)}\frac{\lambda^2}{\pi^4} \\ & + \left( \frac{192 (\log 2n +\gamma +2)+32}{27 \pi ^6 n^2}+O\left(\frac{1}{n^3}\right) \right) \lambda^3 + O(\lambda^4) \text{ for $n>0$},\end{split}
\end{equation}
\begin{equation}
\begin{split}
    \Delta_0^{B} & = 2\Delta_\sigma +\frac{8 \lambda }{\pi ^2} + \frac{\left(14208-928 \pi ^2\right) \lambda ^3}{27 \pi ^6}  +  O(\lambda^4)
    \end{split}
\end{equation}
and 
\begin{equation}
\begin{split}
    \Delta_n^{B} & = 2 \Delta_\sigma + 2n + \frac{64}{3 \pi ^4 n (n+1) (2 n-1) (2 n+1)} \lambda^2 + \\ &    \left( -\frac{384 (\log 2n+\gamma -4)+128}{27 \pi ^6 n^4}+O\left(\frac{1}{n^5}\right)\right)\lambda^3 +  O(\lambda^4) \text{ for $n>0$}.
    \end{split}
\end{equation}

The order $\lambda$ contribution to operators for which $n>0$ vanishes (for both type-A and type-B), as expected, since these are double-trace operators in the theory when $M=1$. 

When $n$ is large,
\begin{equation}
    \lim_{n \to \infty} \gamma^A_n \sim {n^{-2\Delta_\sigma}},
\end{equation}
and
\begin{equation}
    \lim_{n\to\infty} \gamma^B_n \sim \begin{cases}
    {n^{-4\Delta_\psi}} & \lambda \neq 8/5 \\
    {n^{-10/3}} & \lambda =8/5.
    \end{cases}
\end{equation}

\subsection{Leading twist higher-spin operators}
We next present the spectrum of leading-twist higher-spin operators, i.e., those which become almost-conserved currents when $\lambda \to 0$.

The twists of such operators, which are all parity-even, solve an equation of the form 
\begin{equation}
    \det \left( \mathbf K_s^{\text{even}}(\tau) - \mathbf 1 \right) = 0, \label{higher-spin-parity-even-kernel}
\end{equation}
where $\mathbf K_s^{\text{even}}(\tau)$ is given in Appendix \ref{app-integration-kernel}.

\begin{figure}
    \centering
    \includegraphics[width=0.7\textwidth]{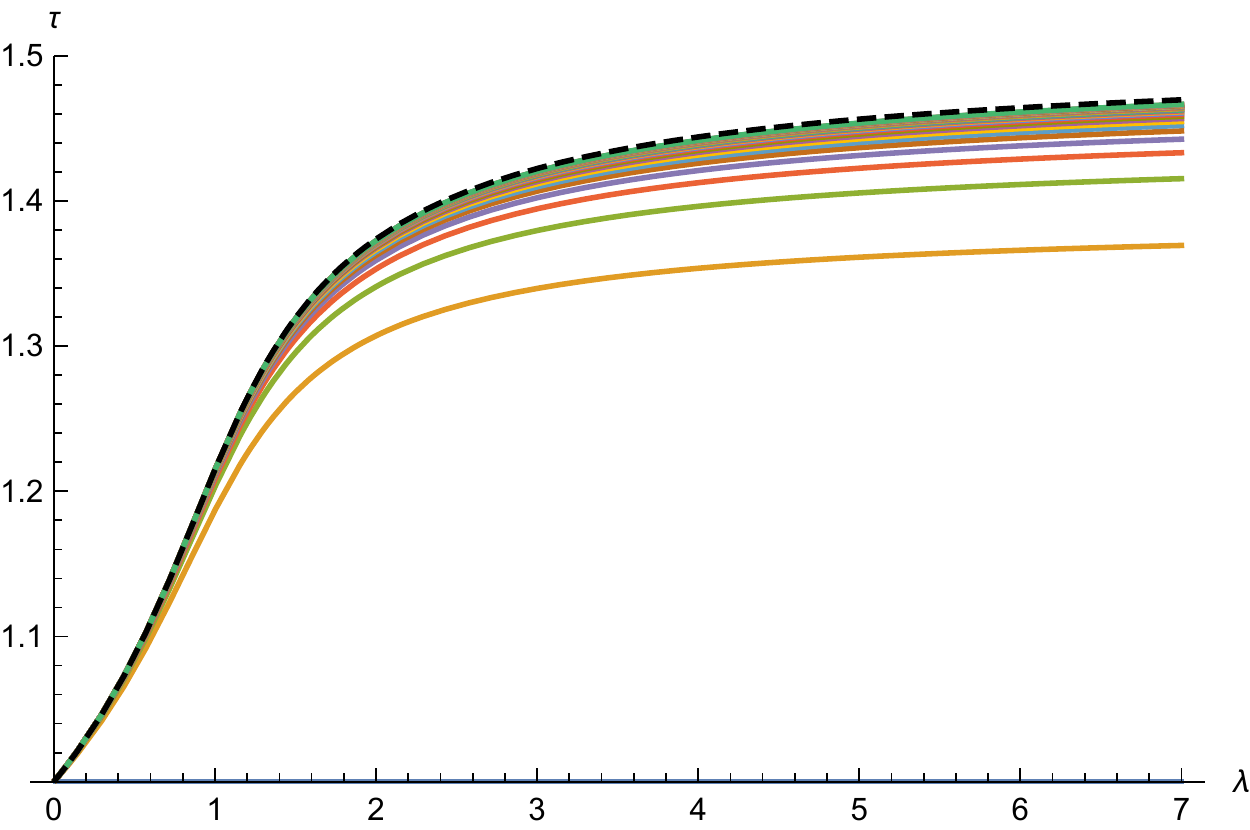}
    \includegraphics[width=0.7\textwidth]{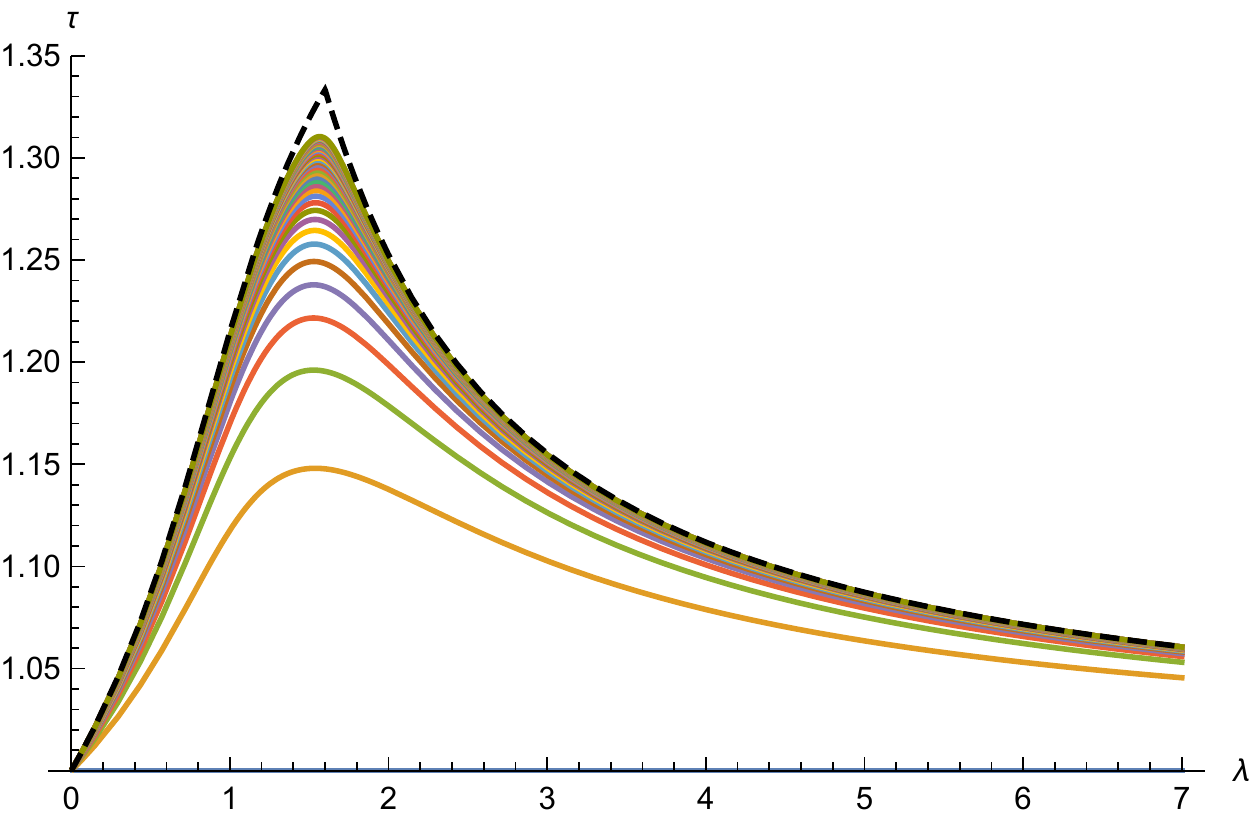}
    \caption{The spectrum of leading-twist operators of spin $s$, for $s$ odd (above) and even (below), is shown for $s=1, \ldots, 80$. As $s\to \infty$, the twists approach the dashed line, and, for even spins, peak near $\lambda=8/5$. \label{higher-spin-spectrum}}
\end{figure}

The resulting twists are shown in Figure \ref{higher-spin-spectrum}. For all $\lambda$, there exists a twist-$1$ spin-$1$ and spin-$2$ operator, corresponding to the conserved $U(1)$ current and stress-tensor.

For odd spins, 
\begin{equation}
    \lim_{s\to \infty} \tau^{\text{min}}_{s} = 2\Delta_\psi-1,
\end{equation}
while for even spins, 
\begin{equation}
    \lim_{s\to \infty} \tau^{\text{min}}_{s} = \begin{cases} 2\Delta_\psi-1 & \lambda \leq 8/5 \\
    2\Delta_\sigma & \lambda>8/5
    \end{cases}.
\end{equation}
Define 
 \begin{equation}
    \gamma_s=\tau_s-\lim_{s\to \infty} \tau^{\text{min}}_s.
\end{equation}
For small $\lambda$, 
\begin{equation}
    \gamma_s = \begin{cases}
    -\frac{4\lambda}{\pi ^2 (2 s-1)}+ \frac{16 \left(4 s (2 s-1)^2 H_{s-1}- (2 s-1)^2(2 s+1) (H_{s-\frac{3}{2}}+ 2\log 2) -2 s \left(16 s^2-22 s+13\right) \right)}{3 \pi ^4 (2 s-1)^3 (2 s+1)}\lambda^2 + O(\lambda^3) & s \text{ even} \\
   -\frac{4\lambda}{\pi ^2 (4 s^2-1)} -\frac{16 \left(\left(1-4 s^2\right)^2 (H_{s-\frac{3}{2}} +2 \log (2)) + 2 \left(20 s^4-13 s^2+6 s+2\right) \right)}{3 \pi ^4 \left(4 s^2-1\right)^3}\lambda^2+ O(\lambda^3) & s \text{ odd}
    \end{cases}.
\end{equation}

For $s$ large and even,
\begin{equation}
    \gamma_s \sim  \begin{cases}
    -\frac{\cos (\pi  \Delta_\psi ) \Gamma \left(\frac{7}{2}-3 \Delta_\psi \right) \Gamma \left(\frac{7}{2}-\Delta_\psi \right) \Gamma \left(2 \Delta_\psi -\frac{3}{2}\right) }{\pi ^{3/2} (4 (\Delta_\psi -2) \Delta_\psi +3) \Gamma (3-4 \Delta_\psi )} s^{1-2 \Delta_\psi } & \lambda<8/5 \\
    -\frac{9 \sqrt{\frac{3}{55} \Gamma \left(\frac{8}{3}\right) \Gamma \left(\frac{14}{3}\right)}}{4\cdot2^{5/6} \pi }s^{-2/3} & \lambda=8/5 \\
    \frac{2 \sin (4 \pi  \Delta_\psi ) \cos (\pi  \Delta_\psi ) \Gamma (3-2 \Delta_\psi ) \Gamma \left(\frac{7}{2}-\Delta_\psi \right) \Gamma \left(2 \Delta_\psi -\frac{1}{2}\right) \Gamma \left(3 \Delta_\psi -\frac{7}{2}\right)}{\pi ^{5/2} (2 \Delta_\psi -1)} s^{1-2 \Delta_\psi } & \lambda>8/5.    \end{cases}
\end{equation}
The asymptotic behavior at $\lambda=8/5$ is unusual -- at $\lambda=8/5$, $\Delta_\sigma = \frac{2}{3}$ is the smallest twist in the theory. 
For $s$ odd, we find:
\begin{equation}
    \gamma_s \sim -\frac{ 2(5-2 \Delta_\psi )^2 \sin^2 (2 \pi  \Delta_\psi ) \Gamma (3-2 \Delta_\psi ) \Gamma (4-2 \Delta_\psi )}{\pi^2(2\Delta_\psi -1)^2} (4s)^{-2\Delta_\sigma}.
\end{equation}
The subleading contribution is $s^{2\Delta_\psi-4}=s^{-(\Delta_\sigma+1)}$, which is numerically close to the leading contribution for $1<\Delta_\psi<5/4$. In the limit $\lambda \to 0$, both the leading and subleading terms approach $s^{-2 + O(\lambda)}$.

The spectrum of both parity-odd and even higher-twist, higher-spin operators are also real for all $\lambda$, and are presented in Appendix \ref{app:higher-twist}.

\subsection{Conformal Regge limit}
\begin{figure}
    \centering
    \includegraphics[width=0.7\textwidth]{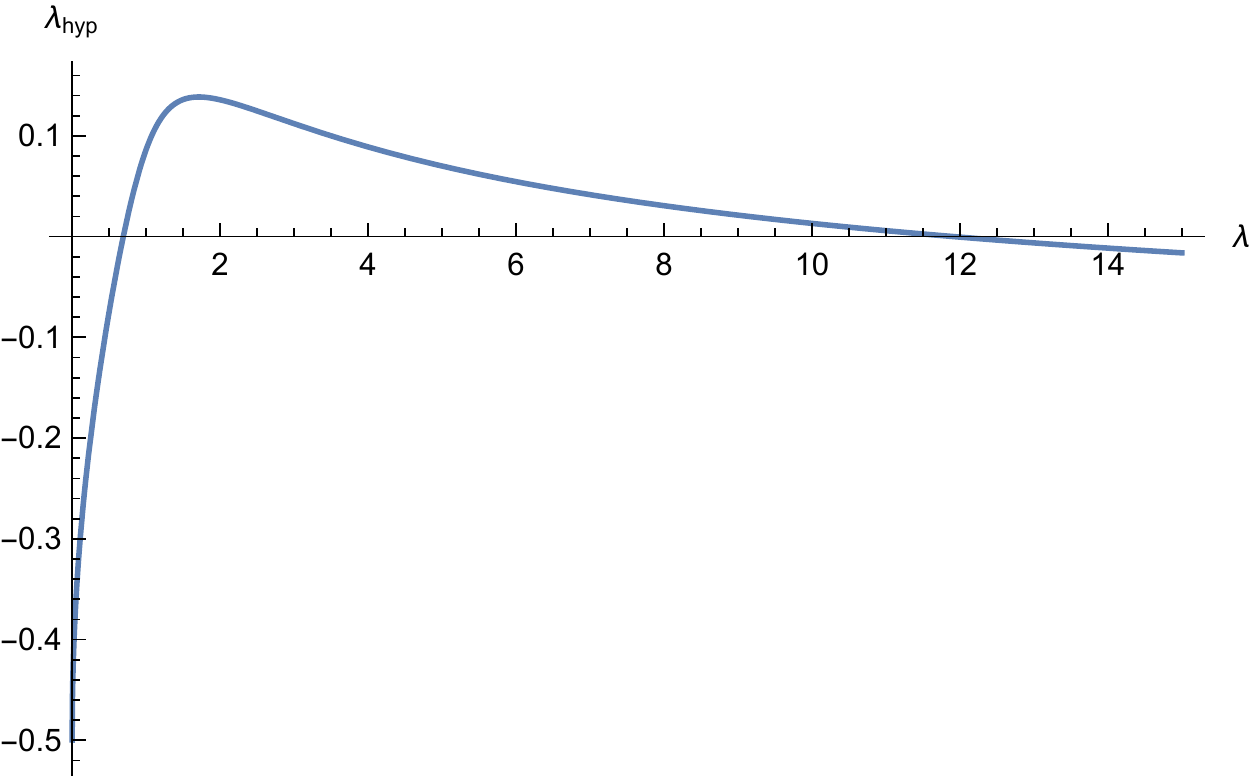}
    \caption{The even-spin hyperbolic Lyapunov exponent $\lambda_{Hyp}=s_*-1$ as a function of $\lambda$. The maximum of $\lambda_{\text{hyp}}=0.139$ is attained at $\lambda=1.71$.}
    \label{fig:lyaponov}
\end{figure}

As explained in \cite{Murugan:2017eto}, if we continue the function $\Delta(s)$ that encodes the scaling dimensions of leading twist operators of spin $s$ to complex spins, the value of $s_*$ that solves $\Delta(s_*)=\frac{3}{2}$
can be used to extract the Lyapunov exponent in a particular hyperbolic geometry via $\lambda_{hyp}=s_*-1$.
The result\footnote{This computation uses the analytically-continued spectrum $\Delta(s)$ of leading-twist, parity-even operators of even spin.} is shown in Figure \ref{fig:lyaponov}, and is similar to \cite{Chang:2021wbx}.

\section{Discussion}
In this paper, we showed that the spectrum of single-trace operators in an IR fixed point defined via a disordered Gross-Neveu Yukawa model  is real for all values of $M/N$. This strongly suggests the model is the first real, nonsupersymmetric CFT in $d>1$ that admits an SYK-like large $N$ limit. Moreover, the theory is a rare example of a solvable, nonsupersymmetric interacting large-$N$ CFT in three dimensions. 

Several questions exist for further research.

When $M/N=0$, the CFT is dual to a higher-spin gauge theory \cite{Sezgin:2003pt}. Is the theory at finite $M/N$ dual to a deformation of this higher-spin gauge theory? 

Our results also suggest that a bifundamental Gross-Neveu Yukawa model, without disorder, could perhaps define a real planar large $N$ CFT in $d=3$.

Disordered models with Yukawa interactions \cite{Wang:2019bpd, Esterlis:2019ola, Chowdhury:2019fcl, Wang:2020dtj, Esterlis:2021eth} and certain coupled SYK/tensor models \cite{Kim:2019upg, Klebanov:2020kck} have been studied as toy models for superconductivity. The model studied here does not exhibit $U(1)$ symmetry breaking, but may perhaps be generalized to one which does.

We hope to return to these questions in the near future.

\begin{acknowledgments}
The author thanks I. Klebanov,  R. Loganayagam, S. Minwalla, M. Rangamani, R. Sinha and S. Sachdev for discussions and comments on a draft of this publication. The author also thanks International Centre for Theoretical Sciences, Bengaluru for hospitality where part of this work was completed. The author is supported in part by DST grants MTR/2018/0010077 and CRG/2021/009137. 
\end{acknowledgments}

\appendix

\section{Solving the gap equation from UV to IR}
\label{numerics}
In this Appendix, we solve the gap equation from weak to strong coupling for $M \ll N$ very small, and $M \gg N$. From these results it seems natural to expect that a flow also exists for intermediate values of $M/N$.

Let $F(p)=1/f(p)$ and $G(p)=-i\slashed{p}/g(p)$.  Note that $[J]=1$ in $d=3$, so we expect these to be functions of the dimensionless quantity $p/J$.

The gap equations are:
\begin{eqnarray}
    f(p) & = & p^2 - J d_\gamma \int \frac{d^dq}{(2\pi)^d} \frac{q^2+q\cdot p}{g(q)g(p+q)} \\
    g(p) & = & p^2  + J\lambda \int \frac{d^dq}{(2\pi)^d} \frac{p\cdot q}{g(q)f(p-q)}.
\end{eqnarray}
We will show these equations have a solution interpolating from the free UV solution to the IR solution at the extreme values of $\lambda$, $\lambda=0$ and $\lambda=\infty$, which are easier to handle than intermediate values of $\lambda$. We have not attempted to numerically solve the gap equation for general $\lambda$.

The $\lambda=0$ solution is trivial, and can be obtained analytically. We have $g^{(\lambda=0)}(p)=p^2$, and $f^{(\lambda=0)}(p)$ is given by
\begin{eqnarray}
    f^{(\lambda=0)}(p) & = & p^2 - J d_\gamma \int \frac{d^dq}{(2\pi)^d} \frac{q^2+q\cdot p}{q^2(p+q)^2} \\
     & = & p^2-d_\gamma J \frac{2^{-d} \pi ^{1-\frac{d}{2}} \csc \left(\frac{\pi  d}{2}\right) \Gamma \left(\frac{d}{2}\right)}{\Gamma (d-1)} p^{d-2} \\
     & \to_{d\to3} & p^2+ \frac{J}{8} p
\end{eqnarray}

\begin{figure}
    \centering
    \includegraphics[width=0.9\textwidth]{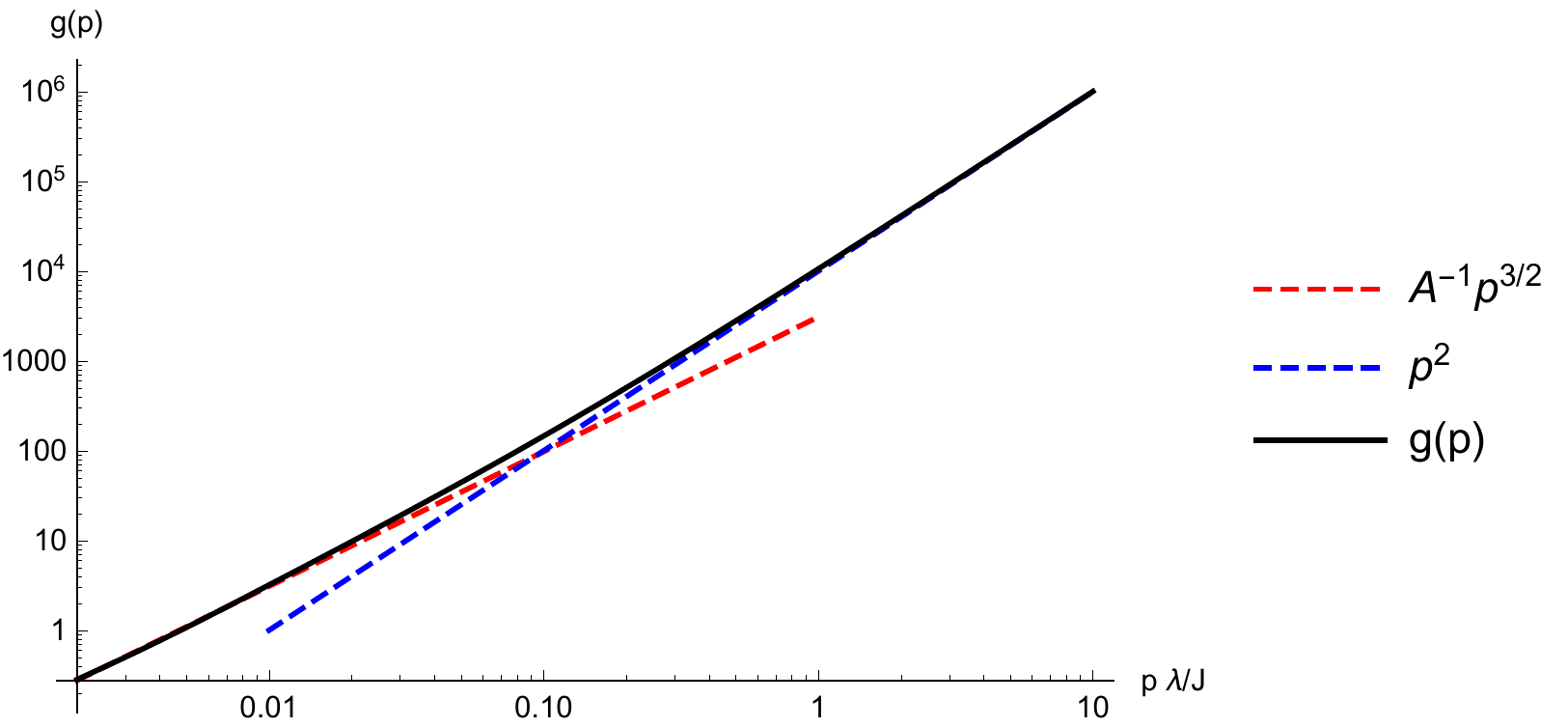}
    \caption{A numerical solution of the gap equation from UV to IR, for the case of $\lambda \to \infty$ is shown above (the solid black line). It interpolates from $g(p) \sim A^{-1}p^{3/2}$ (dashed red line) to $g(p) \sim p^2$ (dashed blue line) as expected.}
    \label{fig:numerical}
\end{figure}

For $\lambda \to \infty$, we must solve the gap equation numerically. Let $\tilde{J}=\lambda J$. Then $f^{(\lambda=\infty)}(p)=p^2$, and $g^{(\lambda=\infty)}(p)$ satisfies:
\begin{equation}
       g^{(\lambda=\infty)}(p)  =  p^2  + \tilde{J} \int \frac{d^3q}{(2\pi)^3} \frac{p\cdot q}{g^{(\lambda=\infty)}(p)(p-q)^2}.
\end{equation}
which we can write as
\begin{equation}
    g^{(\lambda=\infty)}(p)=p^2 + \frac{\tilde{J}}{p(2\pi)^2}\int_0^\infty \frac{q^3 dq}{g^{(\lambda=\infty)}(p)} \mathcal F (q/p), \label{numerical-kernel}
\end{equation}
where 
\begin{equation}
    \mathcal F(q)=\int_{-1}^1 dt \frac{t}{1+q^2-2qt}.
\end{equation}
We solve this equation iteratively, choosing the initial seed function as $g_{\text{seed}}(p) = p^2 + A^{-1}p^{3/2}$, where $A^2=\frac{10 \pi }{3 \tilde{J}}$.
Within a few iterations, this converged to the numerical function plotted in Figure \ref{fig:numerical}.

\section{Details of the computation of the spectrum}
\label{app-integration-kernel}
\subsection{Conformally-invariant three-point functions}
Let $\mathcal O_{\tau,s}^{odd/even}(x,z)$ be a parity-odd/parity-even operator with spin $s$ and twist $\tau=\Delta_{\mathcal O}-s$. There are two allowed forms \cite{GPY} for conformally-invariant three-point correlation functions $$\langle \bar{\lambda}_1 \psi(x_1) \mathcal O_{\tau,s}^{odd/even}(x_3,z_3) \bar{\psi}(x_2) \lambda_2 \rangle,$$ which are:
\begin{equation}
\langle \bar{\lambda}_1 \psi(x_1) \mathcal O_{\tau,s}^{odd}(x_3,z_3) \bar{\psi}(x_2) \lambda_2 \rangle = \frac{1}{x_{31}^\tau x_{12}^{2\Delta_\psi-1-\tau} x_{23}^\tau}  \left(b_1(S_3/P_3)Q_3^s + b_2 (S_1/P_1)P_2 Q_3^{s-1} \right),   \label{parity-odd-form}
\end{equation}
for parity-odd operators, and 
\begin{equation}
\langle \bar{\lambda}_1 \psi(x_1) \mathcal O_{\tau,s}^{even}(x_3,z_3) \bar{\psi}(x_2) \lambda_2 \rangle = \frac{1}{x_{31}^\tau x_{12}^{2\Delta_\psi-1-\tau} x_{23}^\tau}  \left(a_1P_3Q_3^s + a_2 P_1P_2 Q_3^{s-1} \right), \label{parity-even-form}
\end{equation}
for parity-even operators. Here $Q_i$, $P_i$ and $S_i$ are defined in \cite{GPY}, as,
\begin{equation}
    P_3= \bar{\lambda}_1 \frac{\slashed{x}_{12}}{x_{12}^2} \lambda_2,\quad Q_3 = 2z_3 \cdot \left( \frac{x_{32}}{x_{32}^2} - \frac{x_{31}}{x_{31}^2} \right), \quad S_3/P_3 = i \bar{\lambda}_2 \slashed{x}_{23} \slashed{x}_{31} \lambda_1 /(|x_{12}||x_{23}||x_{31}|).
\end{equation}
For $s=0$, the second term in each of the two above expressions should be omitted, and there is only one allowed form for the three-point function.

The correlation function $\langle \sigma(x_1)\sigma(x_2) \mathcal O^{odd}_{\tau,s} \rangle$ vanishes. The only allowed form for a conformally-invariant correlation function $\langle \sigma(x_1)\sigma(x_2) \mathcal O^{even}_{\tau,s}(x_3,z_3) \rangle$ is
\begin{equation}
    \langle \sigma(x_1)\sigma(x_2) \mathcal O^{even}_{\tau,s}(x_3) \rangle= \frac{c_1 Q_3^s}{x_{12}^{2\Delta_\sigma - \tau} x_{23}^\tau x_{31}^{\tau}}. \label{scalar-form}
\end{equation}

It is convenient to take the limit $x_3\cdot z_3=0$ and $x_3 \to \infty$, which can be obtained by a conformal transformation. This makes $Q_3 \to 2 z_3 \cdot \frac{{x}_{21}}{x_{3}^2}$ We also rescale the correlation functions by a factor of $x_3^{2\tau+2s}$. In this limit the allowed forms for the correlation functions become:
\begin{eqnarray}
    v_{a_1} & = &  \bar{\lambda}_1 \slashed{x}_{12} \lambda_2 \frac{(x_{12}\cdot \epsilon)^s}{ |x_{12}|^{-\tau+1+2\Delta_\psi}} \\
    v_{a_2} & = &  \bar{\lambda}_1 \slashed{\epsilon} \lambda_2 \frac{(x_{12}\cdot \epsilon)^{s-1}}{ |x_{12}|^{-\tau-1+2\Delta_\psi}} \\
    v_{b_1} & = &  \bar{\lambda}_1 \lambda_2  \frac{(x_{12}\cdot \epsilon)^s}{|x_{12}|^{2\Delta_\psi - \tau}} \label{vb1}\\
    v_{b_2} & = &  \bar{\lambda}_1 \slashed{\epsilon}\slashed{x}_{12} \lambda_2  \frac{(x_{12} \cdot \epsilon)^{s-1}}{|x_{12}|^{2\Delta_\psi - \tau}} \\
    v_{c_1} & = &  \frac{(x_{12}\cdot \epsilon)^s}{|x_{12}|^{2\Delta_\sigma-\tau}}
\end{eqnarray}
We may prefer to choose a different basis for the parity-odd three-point functions, by defining, 
\begin{equation}v_{b_2}'  =   \bar{\lambda}_1 (\slashed{z}\slashed{x}_{12} - \slashed{x}_{12} \slashed{z})\lambda_2   \frac{(x_{12} \cdot z)^{s-1}}{|x_{12}|^{2\Delta_\psi - \tau}}, \label{vb2prime}\end{equation}
which, it turns out, does not mix with $v_{b_1}$.

\subsection{Integration kernel}
Using the above results, we compute the integration kernel as usual.  

For parity-even spin-$s$ operators, the diagramatic equation in Figure \ref{kernel-figure} of the main text translates into the eigenvalue equation for the coefficients $a_1$, $a_2$ and $c_1$ defined in equations \eqref{parity-even-form} and \eqref{scalar-form}:
\begin{equation} 
\begin{pmatrix} a_1 \\ a_2 \\ c_1 \end{pmatrix} =\mathbf{K}^{\text{even}}_s(\tau) \begin{pmatrix} a_1 \\ a_2 \\ c_1 \end{pmatrix}.
\end{equation}
We must determine the values of $\tau$ for which the $3 \times 3$ matrix $\mathbf{K}^{\text{even}}_s(\tau)$ has an eigenvalue equal to $1$. Similarly, for the spectrum of parity-odd spin-$s$ operators, Figure \ref{kernel-figure} of the main text translates into an eigenvalue equation for a $2\times 2$ matrix $\mathbf{K}^{\text{odd}}_s (\tau)$. For parity-even spin zero operators, Figure \ref{kernel-figure} of the main text translates into an eigenvalue equation for another $2\times 2$ matrix, $\mathbf{K}^{\text{even}}_0 (\tau)$. For  parity odd spin-zero operators, Figure \ref{kernel-figure} of the main text translates into a conventional integration kernel, i.e., a $1 \times 1$ matrix, $K^{\text{odd}}_0 (\tau)$. 

The nine entries of the integration kernel matrix for spin-$s$ parity-even operators are given by
\begin{equation}
\begin{split}
    & \mathbf{K}^{\text{even}}_s (\tau) =  \\ &
    \begin{pmatrix}
    K_{a_1,a_1} [v_{a_1}(y_1,y_2)]/v_{a_1}(x_1,x_2) & K_{a_1,a_2} [v_{a_2}(y_1,y_2)]/v_{a_1}(x_1,x_2) & K_{a_1,c} [v_{c}(y_1,y_2)]/v_{a_1}(x_1,x_2) \\
 K_{a_2,a_1} [v_{a_1}(y_1,y_2)]/v_{a_2}(x_1,x_2) & K_{a_2,a_2} [v_{a_2}(y_1,y_2)]/v_{a_2}(x_1,x_2) & K_{a_2,c} [v_{c}(y_1,y_2)]/v_{a_2}(x_1,x_2) \\
     K_{c,a_1} [v_{a_1}(y_1,y_2)]/v_{c}(x_1,x_2) & K_{c,a_2} [v_{a_2}(y_1,y_2)]/v_{c}(x_1,x_2) & K_{c,c} [v_{c}(y_1,y_2)]/v_{c}(x_1,x_2) 
    \end{pmatrix},
    \end{split} \label{parity-even-higher-spin-kernel}
\end{equation}
for $s>0$. For $s$ odd, the last row and last column of this matrix vanish identically. For $s=0$, the second row and second column of the matrix are not present.

The four entries of the integration kernel matrix for spin-zero parity-even operators are given by
\begin{equation}
\mathbf{K}^{\text{even}}_0 (\tau) = 
    \begin{pmatrix}
    K_{a_1,a_1} [v_{a_1}(y_1,y_2)]/v_{a_1}(x_1,x_2) & K_{a_1,c} [v_{c}(y_1,y_2)]/v_{a_1}(x_1,x_2) \\
     K_{c,a_1} [v_{a_1}(y_1,y_2)]/v_{c}(x_1,x_2) & K_{c,c} [v_{c}(y_1,y_2)]/v_{c}(x_1,x_2) 
    \end{pmatrix}, \label{parity-even-scalar-kernel}
\end{equation}
for $s=0$.

In the above matrices, we have:
\begin{equation}
\begin{split}
    &K_{a_1,a_1}[v_{a_1}(y_1,y_2)] + K_{a_2,a_1}[v_{a_1}(y_1,y_2)] \\ & = J \lambda \int d^d y_1 d^d y_2 \bar{\lambda}_1 G(x_1,y_1) \slashed{y}_{12} G(y_2,x_2) \lambda_2 \frac{(y_{12}\cdot \epsilon)^s}{y_{12}^{-\tau+1+2\Delta_\psi}} F(y_1,y_2),
    \end{split}
\end{equation}
\begin{equation}
    K_{c,a_1}[v_{a_1}(y_1,y_2)] = -J \int d^d y_1 d^d y_2 F(x_1,y_1) F(y_2,x_2) \tr \left(\slashed{y}_{12}G(y_2,y_1) \right) \frac{(y_{12}\cdot \epsilon)^s}{y_{12}^{-\tau+1+2\Delta_\psi}},
\end{equation}
\begin{equation}
\begin{split}
   & K_{a_1,a_2}[v_{a_2}(y_1,y_2)] + K_{a_2,a_2}[v_{a_2}(y_1,y_2)] \\ & = J \lambda \int d^d y_1 d^d y_2 \bar{\lambda}_1 G(x_1,y_1) \slashed{\epsilon} G(y_2,x_2) \lambda_2 \frac{(y_{12}\cdot \epsilon)^{s-1}}{y_{12}^{-\tau-1+2\Delta_\psi}} F(y_1,y_2),
    \end{split}
\end{equation}
\begin{equation}
    K_{c,a_2}[v_{a_2}(y_1,y_2)] = -J \int d^d y_1 d^d y_2 F(x_1,y_1) F(y_2,x_2) \tr \left(\slashed{\epsilon}G(y_2,y_1) \right) \frac{(y_{12}\cdot \epsilon)^{s-1}}{y_{12}^{-\tau-1+2\Delta_\psi}},
\end{equation}
\begin{equation}\begin{split}&
    K_{a_1,c}[v_{c}(y_1,y_2)] + K_{a_2,c}[v_{c}(y_1,y_2)] \\ &= J \lambda \int d^d y_1 d^d y_2 \bar{\lambda}_1 G(x_1,y_1) G(y_1,y_2) G(y_2,x_2) \lambda_2 \frac{(y_{12}\cdot \epsilon)^{s}}{y_{12}^{-\tau+2\Delta_\sigma}},
    \end{split}
\end{equation}
and
\begin{equation}
    K_{c,c}[v_{c}(y_1,y_2)] = 0.
\end{equation}

For parity-odd spin $s$ operators, we have 
\begin{equation}
\mathbf{K}^{\text{odd}}_s (\tau) = 
    \begin{pmatrix}
    K_{b_1,b_1} [v_{b_1}(y_1,y_2)]/v_{b_1}(x_1,x_2) & K_{b_1,b_2} [v_{b_2}(y_1,y_2)]/v_{b_1}(x_1,x_2)  \\
 K_{b_2,b_1} [v_{b_1}(y_1,y_2)]/v_{b_2}(x_1,x_2) & K_{b_2,b_2} [v_{b_2}(y_1,y_2)]/v_{b_2}(x_1,x_2)  \end{pmatrix},
\end{equation}
where,
\begin{equation}
\begin{split} &
    K_{b_1,b_1}[v_{b_1}(y_1,y_2)] + K_{b_2,b_1}[v_{b_1}(y_1,y_2)] \\ & = J \lambda \int d^d y_1 d^d y_2 \bar{\lambda}_1 G(x_1,y_1) G(y_2,x_2) \lambda_2 \frac{(y_{12}\cdot \epsilon)^s}{y_{12}^{-\tau+2\Delta_\psi}} F(y_1,y_2),
    \end{split}
\end{equation}
and
\begin{equation}
\begin{split}
    & K_{b_1,b_2}[v_{b_2}(y_1,y_2)] + K_{b_2,b_2}[v_{b_2}(y_1,y_2)]\\  &= J \lambda \int d^d y_1 d^d y_2 \bar{\lambda}_1 G(x_1,y_1) \slashed{\epsilon} \slashed{y}_{12} G(y_2,x_2) \lambda_2 \frac{(y_{12}\cdot \epsilon)^{s-1}}{y_{12}^{-\tau+2\Delta_\psi}} F(y_1,y_2).
    \end{split}
\end{equation}
$\mathbf{K}^{\text{odd}}_s (\tau)$ is a $2\times2$ matrix for $s>0$. For parity-odd spin $0$ operators, there is no mixing and only one allowed form for the three-point function. The integration kernel $K^{\text{odd}}_0 (\tau)$ is therefore a $1\times 1$ matrix:
\begin{equation}
    K^{\text{odd}}_0 (\tau) =K_{b_1,b_1} [v_{b_1}(y_1,y_2)]/v_{b_1}(x_1,x_2). \label{odd-scalar-kernel}
\end{equation}

Before we proceed, let us point out a technical complication. Unlike the usual SYK model, or the higher-dimensional bosonic variants, we find that there are multiple allowed forms for the three-point correlation functions. Like mixing, as discussed in \cite{Murugan:2017eto, GKPPT, Chang:2021wbx}, this means the integration kernel is a matrix -- and the allowed scaling dimensions are those for which is an 1 eigenvalue of the matrix. When two operators mix we expect two linearly independent mixtures to exist that have well-defined scaling dimensions. This is reflected in the fact that the calculation of the integration kernel involves diagonalizing a two-by-two matrix and eventually leads to two sets of spectra, both of which are physical. However, suppose there are multiple allowed forms for the three-point correlation function -- for simplicity, assume that there are $n$ allowed forms for the three point function and no mixing. In the actual IR fixed point, only one three-point function would be realized, which corresponds to one of the eigenvectors of an $n\times n$ integration-kernel matrix, with eigenvalue $1$. The other eigenvectors, and their associated eigenvalues, would be spurious. We therefore determine $n$ candidate scaling dimensions for each operator, without identifying which of these scaling dimensions is actually realized. Luckily, we are able to resolve this ambiguity by assuming the spectrum of the theory is a continuous function of $\lambda$. This complication does not affect spin-zero operators.

\subsection{Numerically solving for the allowed scaling dimensions}
\label{app:numerical-spectrum-details}
To determine the allowed scaling dimensions of parity-odd scalars, we must determine the values of $\tilde{\Delta}$ for which $K_0^{\text{odd}}(\tilde \Delta)=1$. 
This gives rise to
\begin{equation}
    -\frac{\Gamma \left(\frac{1}{2} (d-2 \Delta_\psi +1)\right) \Gamma \left(d-\Delta_\psi +\frac{1}{2}\right) \Gamma \left(\Delta_\psi -\frac{\tilde{\Delta} }{2}\right) \Gamma \left(-\frac{d}{2}+s+\Delta_\psi +\frac{\tilde{\Delta} }{2}\right)}{\Gamma \left(\Delta_\psi +\frac{1}{2}\right) \Gamma \left(-\frac{d}{2}+\Delta_\psi +\frac{1}{2}\right) \Gamma \left(d-\Delta_\psi -\frac{\tilde{\Delta} }{2}\right) \Gamma \left(\frac{1}{2} (d+2 s-2 \Delta_\psi +\tilde{\Delta} )\right)}=1.
\end{equation}

\begin{figure}
    \centering
    \includegraphics[width=0.7\textwidth]{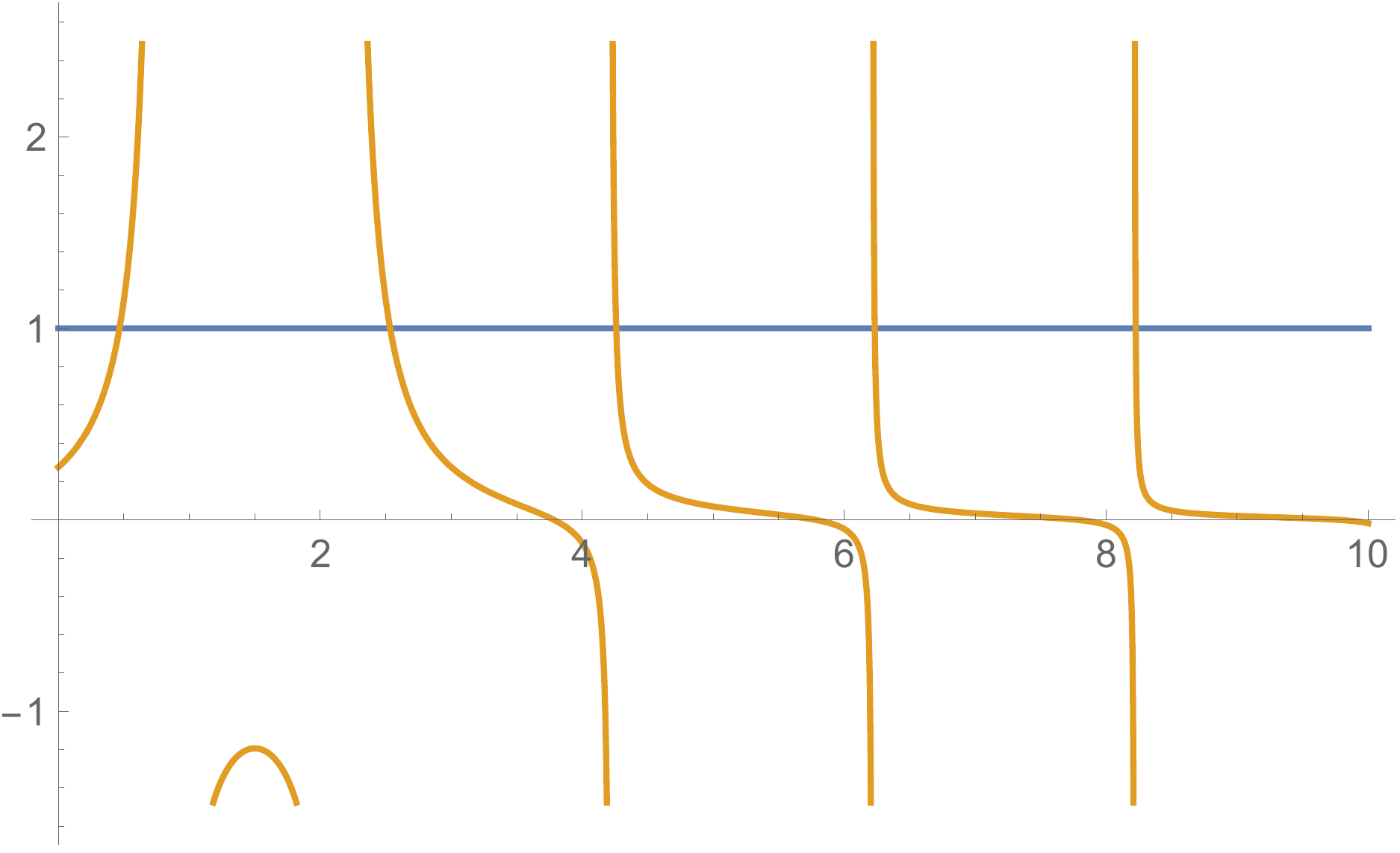}
    \caption{Computing the spectrum of spin-zero parity-odd scalars in $d=3$ for $\lambda=1$. For other values of $\lambda$, the calculation is very similar, and we find the lowest twist parity-odd scalar has real scaling dimension for all $\lambda$, in contrast to other nonsupersymmetric melonic theories, such as \cite{Prakash:2017hwq}.}
    \label{fig:spectrum-0}
\end{figure}

\begin{figure}
    \centering
    \includegraphics[width=0.7\textwidth]{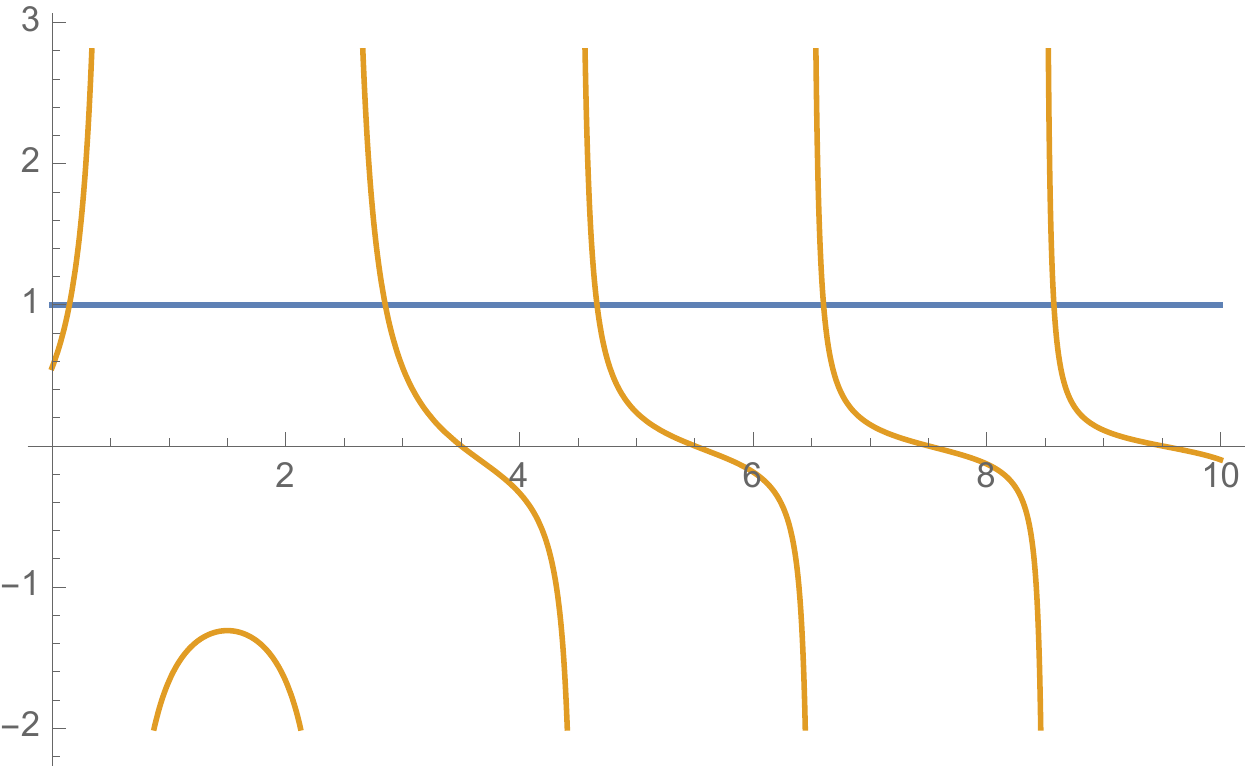}
    \caption{Computing the spectrum of spin-zero parity-odd scalars in $d=3$ for $\lambda=\infty$. }
    \label{fig:spectrum-0-infty}
\end{figure}

We carry out this computation numerically for $d=3$, as illustrated in Figures \ref{fig:spectrum-0} and \ref{fig:spectrum-0-infty} for $\lambda=1$ and $\lambda=\infty$ respectively. These figures illustrate that the spectrum is real for these values of $\lambda$, and we established analytically that the spectrum is real for small $\lambda$. We also determined the spectrum for numerous intermediate values of $\lambda$ and found that it is real as well.

Let us discuss a potential concern regarding the spectrum. When $\lambda=0$ the disordered spectrum is that of the large $N$ GN model. One might also expect that the $O(\lambda^1)$ correction to the disordered spectrum matches the $O(1/N)$ correction to the spectrum of the GN model, but this is not quite true. The $1/N$ contribution to scaling dimension of the single-trace scalar primary in the GN model is $\tilde{\Delta}_0'=1-\frac{16}{3\pi^2}\frac{1}{N}$, which is the shadow of the value quoted for $\tilde{\Delta}_0$ above when $\lambda$ is replaced by $1/N$. The reason for this discrepancy is as follows. When $M=1$, in the GN model, the scalar field $\sigma$ can be integrated out, and its equation of motion sets $\bar{\psi}\psi=0$, effectively replacing the scalar primary with $\sigma$. When one computes the $1/N$ anomalous dimension of $\sigma$ in the GN model, using e.g., Feynman diagrams, one obtains the correction $-\frac{16}{3\pi^2}\frac{1}{N}$, leading to the value $\Delta_0'$ quoted above. However, in the disordered theory with $M>1$, we cannot integrate out $\sigma$, so we cannot set $\bar{\psi}\psi$ to zero --  its scaling dimension becomes meaningful, and is equal to $\tilde{\Delta}_0$ computed above for any $\lambda>0$.  The order $\lambda$ correction to the scaling dimension of $\sigma^a$ also does not match the $1/N$ correction to the scaling dimension of $\sigma$ in the GN model -- this is because the Feynman diagrams that contribute to the $1/N$ anomalous dimension of $\sigma$ in the GN model include a non-melonic diagram that is suppressed by $1/N$ rather than $M/N$ in the disordered theory.

\section{Higher-twist spin-$s$ operators in $d=3$}
\label{app:higher-twist}

\subsection{Parity-even higher-spin operators}

\begin{figure}
    \centering
    \includegraphics[width=0.7\textwidth]{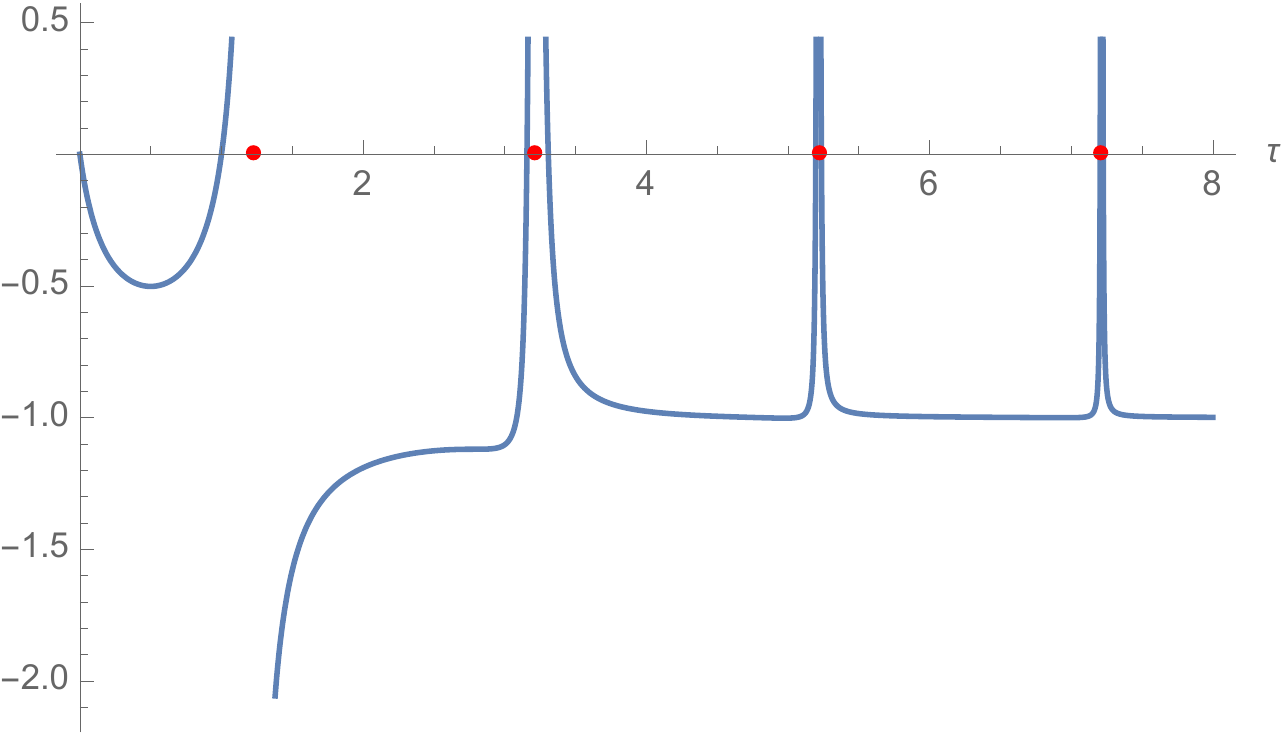}
    \includegraphics[width=0.7\textwidth]{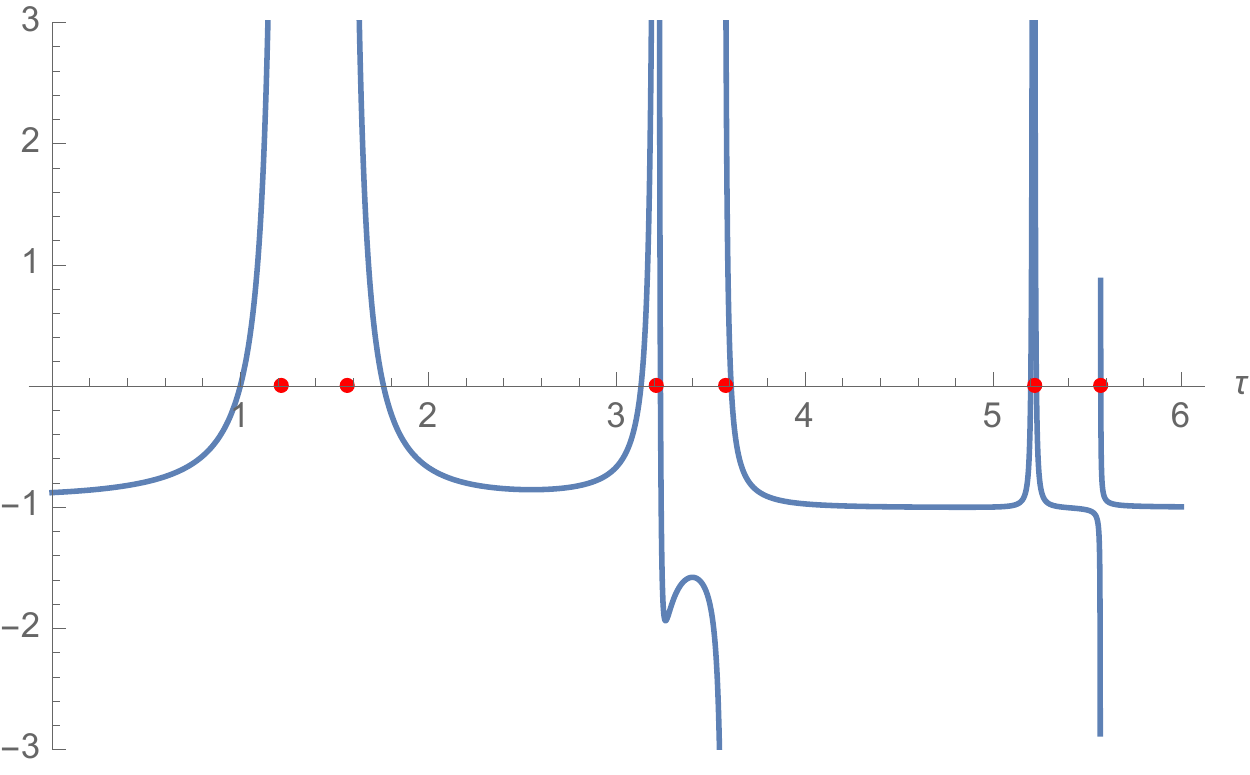}
     \caption{Determining the spectrum of parity-even single-trace spin-$1$ (above) and spin-$2$ (below) operators for $\lambda=1$. Red dots denote asymptotic values of twists, approached as $n \to \infty$. }
    \label{fig:s12}
\end{figure}

The calculation leading to the numerical spectrum for $s=1,2$ and $\lambda=1$ is shown in Figure \ref{fig:s12}. The plots for higher-spin operators are similar, except that the leading twist is no longer at $\tau=1$.

\begin{figure}
    \centering
    \includegraphics{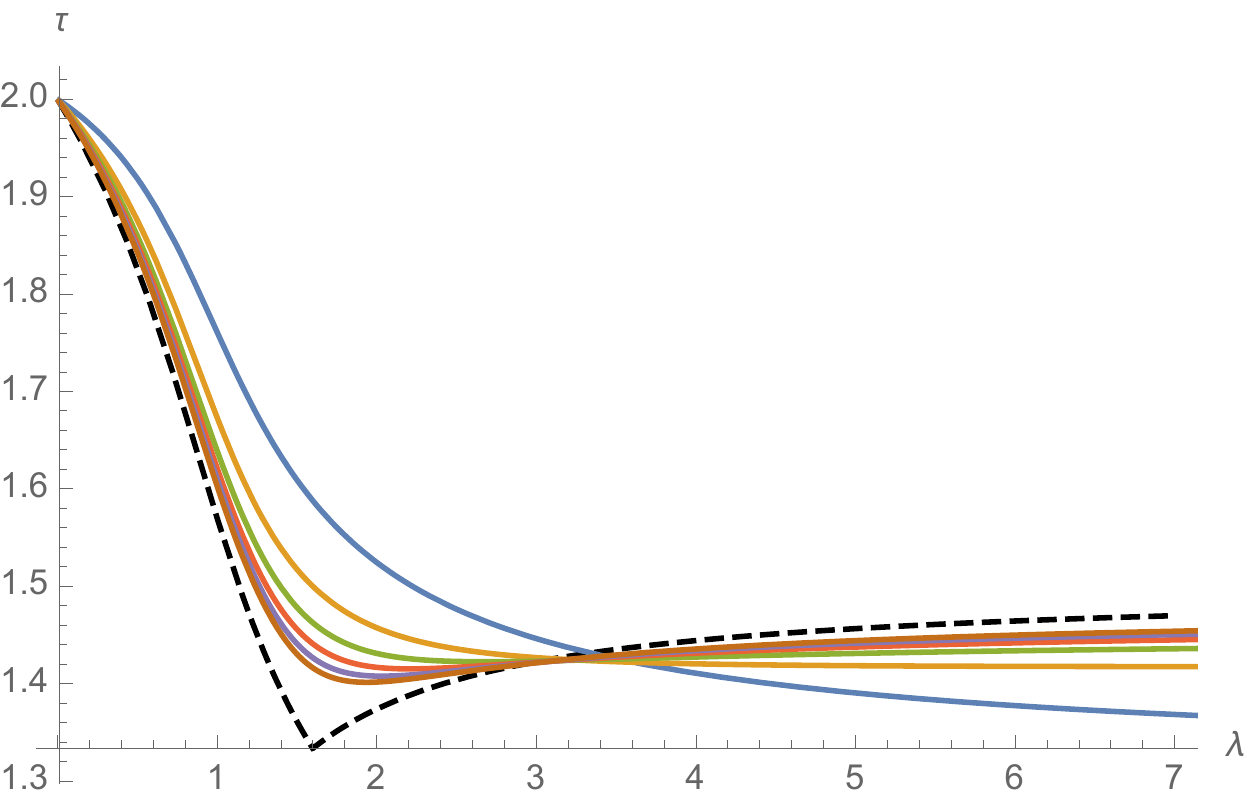}
    \caption{The anomalous dimensions of the even spin higher-spin operators of sub-leading twist. The dashed-line is the asymptotic value approached as $s\to \infty$. For sufficiently large $\lambda$ these are monotonically increasing functions of $s$.}
    \label{fig:subleading}
\end{figure}

From Figure \ref{fig:s12}, we see that, for operators of both leading and subleading twist, there is no ambiguity in the spectrum. The spectrum of operators of sub-leading twists is shown in Figure \ref{fig:subleading} as a function of $\lambda$. 

However, as discussed in Appendix \ref{app-integration-kernel}, there is an ambiguity in determining the higher-twist higher-spin operators. Because there are two allowed forms for the correlation function involving two fermions and a spin-s operator, we find two values of $\tau^A_{s,n}$  for which $K=1$, for each $s>0$ and $n>0$. These approach $2\Delta_\psi-1$ from above and below as $n \to \infty$ or $s \to \infty$, so we denote them as $\tau^{A+}_{s,n}$ and $\tau^{A-}_{s,n}$ respectively. Only one of $\tau^{A\pm}_{s,n}$ is physical. Because there is only one allowed form for $\langle \sigma \sigma J_s \rangle$, no such ambiguity is present for $\tau^B_{s,n}$. This can be seen in in Figure \ref{fig:s12}.

We resolve this ambiguity by demanding anomalous dimensions are continuous functions of $\lambda$.  We find that the functions 
\begin{equation}
    \tau^{(1)}_{s,n}(\lambda) = \begin{cases} \tau^{A-}_{s,n}(\lambda) \lambda< \lambda_* \\
    \tau^{B}_{s,n}(\lambda) & \lambda>\lambda_*
    \end{cases}
\end{equation}
and
\begin{equation}
    \tau^{(2)}_{s,n}(\lambda) = \begin{cases} \tau^{B}_{s,n}(\lambda) \lambda< \lambda_* \\
    \tau^{A+}_{s,n}(\lambda) & \lambda>\lambda_*
    \end{cases}
\end{equation}
are continuous functions of $\lambda$. Therefore \begin{equation}
    \tau^A_{n,s}=
    \begin{cases}
    \tau^{A-}_{n,s} & \lambda<\lambda* \\
     \tau^{A+}_{n,s} & \lambda>\lambda*. 
     \end{cases}
\end{equation} 

\subsection{Parity-odd higher-spin operators}
The same ambiguity arises for parity odd higher-spin operators. In this case, it is possible to explicitly choose a basis -- formed by $v_{b_1}$ and $v_{b_2}'$ in equations \eqref{vb1} and \eqref{vb2prime} -- for correlation functions that diagonalizes the $2\times 2$ integration kernel matrix for all $\Delta_\psi$, $\tau$ and $s$. There are therefore two integration kernels
\begin{equation}
    \tilde{K}^{+}_s(\tilde{\tau}) = -\frac{\Gamma \left(\frac{1}{2} (d-2 \Delta_\psi +1)\right) \Gamma \left(d-\Delta_\psi +\frac{1}{2}\right) \Gamma \left(\Delta_\psi -\frac{\tau }{2}\right) \Gamma \left(-\frac{d}{2}+s+\Delta_\psi +\frac{\tau }{2}\right)}{\Gamma \left(\Delta_\psi +\frac{1}{2}\right) \Gamma \left(-\frac{d}{2}+\Delta_\psi +\frac{1}{2}\right) \Gamma \left(d-\Delta_\psi -\frac{\tau }{2}\right) \Gamma \left(\frac{1}{2} (d+2 s-2 \Delta_\psi +\tau )\right)},
\end{equation}
and
\begin{equation}
    \tilde{K}^-_s(\tilde{\tau}) = \frac{2 \Gamma \left(\frac{1}{2} (d-2 \Delta_\psi +1)\right) \Gamma \left(d-\Delta_\psi +\frac{1}{2}\right) \Gamma \left(\Delta_\psi -\frac{\tau }{2}\right) \Gamma \left(-\frac{d}{2}+s+\Delta_\psi +\frac{\tau }{2}\right)}{\Gamma \left(\Delta_\psi +\frac{1}{2}\right) \Gamma \left(-\frac{d}{2}+\Delta_\psi +\frac{1}{2}\right) \Gamma \left(d-\Delta_\psi -\frac{\tau }{2}\right) \Gamma \left(\frac{1}{2} (d+2 s-2 \Delta_\psi +\tau )\right)}.
\end{equation}
The eigenvector corresponding to  $\tilde{K}^+_{s}$ is $v_{b_1}$, is also valid for $s=0$, while the eigenvector corresponding to $\tilde{K}^-_{s}$ is $v_{b_2}'$ and requires $s \geq 1$.

The equations $\tilde{K}^{\pm}_s(\tilde{\tau})=1$ have solutions $\tilde{\tau}^+_{n,s}$ and $\tilde{\tau}^-_{n,s}$ for each integer $n \geq 0$. $\tilde{\tau}^+_{n,s}$ and $\tilde{\tau}^-_{n,s}$  approach $2\Delta_\psi+2n$ from above and below, respectively, as $n \to \infty$ or $s \to \infty$. 

When we solve these equations numerically, we find that $\tilde \tau^-_{1,0}(\lambda)$ becomes complex near $\lambda=6.375$, while $\tilde \tau^+_{s,0}$ remains real for all $\lambda$. Hence, it is important to know whether  $\tilde \tau^+_{s,n}$ or $\tilde \tau^-_{s,n}$ is realized in the IR limit. 

We carried out a perturbative Feynman diagram calculation of the anomalous dimension of $\bar{\psi} \overleftrightarrow{\partial} (\cdot z)^s \psi$  to first order in $\lambda$ for odd $s$. This is very similar to calculating $1/N$ corrections to the GN vector model, we use the IR propagators at $\lambda=0$: $F(p)=\frac{8\lambda}{J p}$ and $G(p)=G_0(p)$, and, keep track of index contractions arising from disorder-averaging -- which means we keep only melonic diagrams. There is only one diagram that contributes for odd $s$ (other than the self-energy of the fermion) which is given by:
\begin{equation}
    -\int \frac{d^3q}{(2\pi)^3} \frac{8 \lambda}{q (p+q)^2} ((p+q)\cdot z)^s \to -\frac{4}{(2s+1)\pi^2} \lambda \log \Lambda.
\end{equation}
We find that $\tilde{\tau}_{s,0}=2\Delta_\psi+\frac{4}{(2s+1)\pi^2}\lambda+O(\lambda^2)$, for odd $s$, which matches $\tilde \tau^+_{s,0}$ obtained from solving $\tilde{K}^+_s=1$. Assuming anomalous dimensions are continuous functions of $\lambda$, we conclude that $\tilde{\tau}^+_{s,0}$ are the physically realized twists for the leading twist parity-odd higher spin operators, and our spectrum is free from any operators of complex scaling dimension for all $\lambda$. If we assume $\tilde{\tau}_{s,n}$ are monotonic functions of $n$, this also implies that $\tau^+_{s,n}$ are the physically realized twists for $n>0$ -- to confirm this expectation would require an order $\lambda^2$ perturbative calculation. However, for $n>0$, both $\tau^+_{s,n}$ and $\tau^-_{s,n}$ are real so distinguishing between them is not essential for the main conclusion of the paper.

For small $\lambda$, we find that
\begin{equation}\begin{split}
  &  \tilde \gamma_{s,0} = \frac{4}{\pi ^2 (2 s+1)}\lambda \\ & + \frac{16 \left(9 (2 s+1)^2 H_{s-\frac{1}{2}}+80 s^3+12 s^2 (13+\log (64))+12 s (8+\log (64))-8+9 \log (4)\right)}{27 \pi ^4 (2 s+1)^3} \lambda^2 + O(\lambda^3)
    \end{split}
\end{equation}
and
\begin{equation} \begin{split}
    &\tilde \gamma_{s,n}  = \frac{16}{3 \pi ^4 n (2 n+2 s+1)} \lambda^2 \\
    & + \frac{8 \left(6 n (2 n+2 s+1) \left(H_{n+s-\frac{1}{2}}+H_{n-1}+\log (4)\right)+26 n (2 n+2 s+1)+9\right)}{27 \pi ^6 n^2 \left(n+s+\frac{1}{2}\right)^2}\lambda^3 + O(\lambda^4) \\
    &\sim_{n \to \infty} \frac{8}{3 \pi ^4 n^2} \lambda^2+\frac{32 (6 \log (n)+6 \gamma +13+3 \log (4))}{27 \pi ^6 n^2} \lambda^3 + O(\lambda^3) 
    \\
    &\sim_{s \to \infty} \frac{8}{3 \pi ^4 n s} \lambda^2+ \frac{32 \left(3 H_{n-1}+3 \log (s)+3 \gamma +13+3 \log (4)\right)}{27 \pi ^6 n s} \lambda^3 + O(\lambda^4).
    \end{split}
\end{equation}

At large $s$, we find,
\begin{equation}
    \tilde \gamma_{s,n} \sim \frac{2^{4 \Delta_\psi -5} (2 \Delta_\psi -5) \sin (2 \pi  \Delta_\psi ) \Gamma (4-2 \Delta_\psi ) }{\pi  (1-2 \Delta_\psi)} s^{-\Delta_\sigma},
\end{equation}
and at large $n$,
\begin{equation}
    \tilde \gamma_{s,n} \sim \frac{2^{4 \Delta_\psi -5} (5-2 \Delta_\psi) \sin ^2(2 \pi  \Delta_\psi ) \Gamma (3-2 \Delta_\psi ) \Gamma (4-2 \Delta_\psi ) }{\pi ^2 (2 \Delta_\psi -1)}n^{-2\Delta_\sigma}.
\end{equation}

\section{Spectrum in $d=4-\epsilon$}
\label{other-dimensions}

\label{app:epsilon}
We also solve the spectrum of single-trace bilinears in the theory in $d=4-\epsilon$, which allows us to determine analytic expressions for various quantities in terms of $\lambda$.  In this section, it is convenient to absorb $d_\gamma$ into the definition of $\lambda$ by defining $\hat{\lambda}=\frac{2 M}{d_\gamma N}$; alternatively, one can set $d_\gamma=2$ and then $\hat\lambda=\lambda$. 

\subsection{The gap equation}
We first solve the gap equation. The allowed range for $\Delta_\psi$ is $\frac{3}{2}-\frac{\epsilon}{2} \leq \Delta_\psi \leq \frac{3}{2} -\frac{\epsilon}{4}$. We find:
\begin{equation}
    \Delta_\psi= \frac{3}{2}-\frac{(\hat{\lambda}+2) \epsilon }{4 (\hat{\lambda}+1)}+\frac{\hat{\lambda} (5 \hat{\lambda}-6) \epsilon ^2}{32 (\hat{\lambda}+1)^3}+O\left(\epsilon ^3\right),
\end{equation}
and
\begin{equation}
    \Delta_\sigma = 1-\frac{\hat{\lambda} \epsilon }{2 (\hat{\lambda}+1)}-\frac{\hat{\lambda} (5 \hat{\lambda}-6) \epsilon ^2}{16 (\hat{\lambda}+1)^3}+O\left(\epsilon ^3\right).
\end{equation}
The critical value, $\hat{\lambda}_*$, of $\hat \lambda$ at which $2\Delta_\psi(\hat\lambda_*)-1= 2\Delta_\sigma(\hat\lambda_*)$ is given by ${\hat\lambda}_* = 2-\frac{\epsilon}{3}-\frac{\epsilon^2}{18}+O(\epsilon^3)$.

\subsection{Parity-odd scalars}
Solving for the spectrum of parity-odd scalars, we find the lowest scalar has scaling dimension given by:
\begin{equation}
    \tilde{\Delta}_0 = 3+\frac{(\hat\lambda -2) \epsilon }{2 (\hat\lambda +1)}-\frac{\hat\lambda\left(14 \hat\lambda ^2-5 \hat\lambda+14 \right) \epsilon ^2}{16 (\hat\lambda +1)^3}+\frac{\hat\lambda  \left(32 \hat\lambda ^4-57 \hat\lambda ^3+143 \hat\lambda ^2-153 \hat\lambda -22\right) \epsilon ^3}{64 (\hat\lambda +1)^5}+O\left(\epsilon ^4\right).
\end{equation}
The higher-twist parity odd scalars have dimension:
\begin{equation}\begin{split}
\tilde{\Delta}_n  = & 2\Delta_\psi + 2n + \frac{8\epsilon ^2 \hat\lambda^2    (\hat\lambda +1)}{16 (\hat\lambda +1)^3 n (n+1)}
\\
& + \frac{ \epsilon ^3  \hat\lambda ^2  \left(4\hat\lambda (\hat\lambda +1) \left( 2 n (n+1) H_{n-1}+1\right)+\hat\lambda ^2 \left(-7 n^2+n+4\right)+\hat\lambda  (n^2+13n+4)-2 n (7 n+5)\right)}{16 (\hat\lambda +1)^4n^2 (n+1)^2} \\ & + O(\epsilon^4) \end{split}
\end{equation}

The anomalous dimensions for large $n$ are
\begin{equation}
\begin{split}
    \tilde{\gamma}_{0,n} \sim & \frac{\hat\lambda ^2}{2 (\hat\lambda +1)^2 n^2} \epsilon^2 + \frac{\hat\lambda ^2 (\hat\lambda  (-7 \hat\lambda +8 \gamma  (\hat\lambda +1)+1)+8 \hat\lambda  (\hat\lambda +1) \log n-14)}{16 (\hat\lambda +1)^4 n^2} \epsilon^3 + O(\epsilon^4),
\end{split}    
\end{equation}
which is consistent with 
$\tilde\gamma_{0,n} \sim \frac{1}{n^{2\Delta_\sigma}}.$

\subsection{Parity-even scalars}
We next compute the parity-even scalar spectrum. We find for the lowest twist parity-even scalars:
\begin{eqnarray}
    \Delta^A_0 & = & 4-\epsilon \\
    \Delta^B_0 & = & 2-\frac{\left(\hat\lambda-\sqrt{\hat\lambda ^2+14 \hat\lambda +1} +1\right) \epsilon }{2 \hat\lambda +2}-\frac{\hat\lambda  \left(35 \hat\lambda ^2-5 \hat\lambda +26\right) \epsilon ^2}{8 \left((\hat\lambda +1)^3 \sqrt{\hat\lambda ^2+14 \hat\lambda +1}\right)} \nonumber  \\ && +\frac{3 \hat\lambda  \left(25 \hat\lambda ^6+136 \hat\lambda ^5+213 \hat\lambda ^4+3427 \hat\lambda ^3+336 \hat\lambda ^2-83 \hat\lambda +2\right) \epsilon ^3}{32 (\hat\lambda +1)^5 \left(\hat\lambda ^2+14 \hat\lambda +1\right)^{3/2}}+O\left(\epsilon ^4\right)
\end{eqnarray}

The anomalous dimensions of type-A higher twist parity-even scalars are,
\begin{equation}
    \begin{split}
        \gamma^A_{0,n} = & \frac{\hat\lambda ^2 \epsilon ^2}{2 (\hat\lambda +1)^2 n (n+2)} + \\
        & \frac{\hat\lambda ^2 \epsilon ^3}{16 (\hat\lambda -2) (\hat\lambda +1)^4 n^2 (n+1)^2 (n+2)^2} \bigg(8 \hat\lambda  \left(\hat\lambda-2\right)(\hat\lambda+1) (n+1)^2 (n+2) n H_{n-1} \\ & +4 \left(3 \hat\lambda ^3+\hat\lambda ^2+6 \hat\lambda +8\right) 
        + \left(-7 \hat\lambda ^3+15 \hat\lambda ^2-16 \hat\lambda +28\right) n^4 \\ & -4 \left(4 \hat\lambda ^3-13 \hat\lambda ^2+23 \hat\lambda -26\right) n^3+\left(5 \hat\lambda ^3+43 \hat\lambda ^2-168 \hat\lambda +124\right) n^2 \\ & 
        +2 \left(13 \hat\lambda ^3-3 \hat\lambda ^2-58 \hat\lambda +24\right) n\bigg) + O(\epsilon^4) \\
        & \sim_{n \to \infty} \epsilon ^2 \left(\frac{\hat\lambda ^2}{2 (\hat\lambda +1)^2 n^2}\right)  + \epsilon ^3 \left(\frac{\hat\lambda ^2 ( \hat\lambda -7 \hat\lambda^2 +8 \hat\lambda  (\hat\lambda +1) (\log n+\gamma)-14)}{16 (\hat\lambda +1)^4 n^2}\right)+O\left(\epsilon ^4\right),
    \end{split}
\end{equation}
which is consistent with $\gamma^A_{0,n} \sim n^{-2\Delta_\sigma}$ at large $n$. At finite $n$, the expression is slightly different when $\hat\lambda=\hat\lambda_*$.

For type-B scalars, the $n=1$ operator needs to be handled separately:
\begin{eqnarray}
    \gamma^B_{0,1} &=& \frac{\hat\lambda }{\hat\lambda +1} \epsilon-\frac{\hat\lambda  \left(\hat\lambda ^2+8 \hat\lambda +18\right)}{8 (\hat\lambda +1)^3}\epsilon^2+O(\epsilon^3) 
\end{eqnarray}

For $\hat\lambda \neq \hat\lambda_*$, the type-B anomalous dimensions for $n>1$ are:
\begin{equation}
    \gamma^B_{0,n} = -\frac{2\hat\lambda \left(\hat\lambda+2 \right)}{(\hat\lambda -2) (\hat\lambda +1)^3 n^2 \left(n^2-1\right)^2}\epsilon^3+O(\epsilon^4),
\end{equation}
 consistent with $\gamma^B_{0,n} \sim n^{-4\Delta_\psi}$.
 
For $\hat\lambda=\hat\lambda_*=2-\epsilon/3-\epsilon^2/18$ the type-B anomalous dimensions for $n>1$ are:
\begin{equation}
    \begin{split}\gamma^B_{0,n}(\hat\lambda_*) &=  \frac{n-\sqrt{n^2+8}}{9n (n+1)(n-1)}\epsilon^2 +O(\epsilon^3) \\
    &\sim_{n \to \infty} -\frac{4}{9 n^4}\epsilon^2 -\frac{2 (4 \log n+4 \gamma -7)}{27 n^4}\epsilon^3+ O(\epsilon^4),
    \end{split}
\end{equation}
 consistent with $\gamma^B_{0,n}(\hat\lambda_*) \sim n^{-(2\Delta_\psi+1)}$.
 
For type-A scalars, at $\hat\lambda=\hat\lambda_*$, the anomalous dimensions at finite $n$ also differ from the results at generic values of $\hat\lambda$:
 \begin{equation}
     \gamma^A(\hat\lambda_*)_{0,n} = \frac{n+\sqrt{(n+1)^2+8}+1}{9n(n+2)(n+3)}\epsilon + O(\epsilon)^2,
 \end{equation}
 but this does not affect the large $n$ asymptotics.
\subsection{Leading twist higher-spin operators}
Let us compute the leading twist higher-spin operators, which are parity-even.

\begin{figure}
    \centering
    \includegraphics[width=0.7\textwidth]{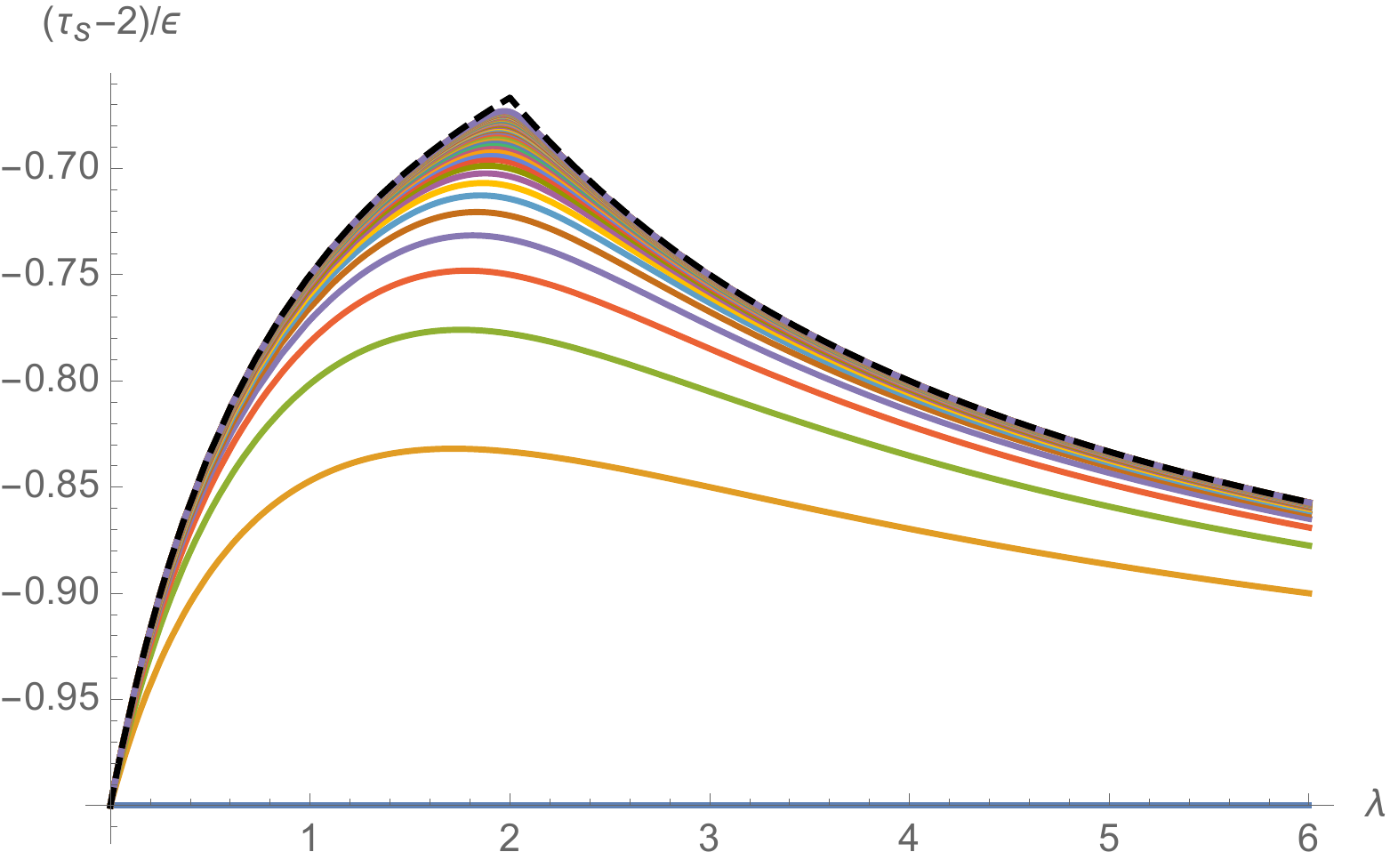}
    \caption{The $O(\epsilon)$ contribution to the twist of leading-twist higher-spin operators in $d=4-\epsilon$, given in equation \eqref{epsilon-expansion-higher-spins} is plotted for even spins $2$, $4$, $\ldots$, $100$. The dashed line is the asymptotic behavior as $s\to \infty$.}
    \label{fig:epsilon-expansion-higher-spins}
\end{figure}

For odd $s$, the leading-twist of a spin $s$ operator with $s \geq 1$, is given by:
\begin{equation}
    \begin{split}   \tau_{s,\text{min}} & = 2\Delta_\psi -1 -\frac{\hat\lambda }{(\hat\lambda +1) \left(s^2+s\right)} \epsilon +\\ &\frac{\hat\lambda   \left(-4 \hat\lambda  (\hat\lambda +1) ((s+1)^2 s^2 H_{s-1}+1)+(7 \hat\lambda^2 +8\hat\lambda+12) s^4+2 \left(3 \hat\lambda ^2+8\right) s^3-\hat\lambda  (\hat\lambda +12) s^2-4 (\hat\lambda +1) (2 \hat\lambda +1) s\right)}{8 (\hat\lambda +1)^3 s^3 (s+1)^3}\epsilon ^2 \\ & +O\left(\epsilon ^3\right) \\
        & =  2-\frac{\epsilon  \left(\hat\lambda +\frac{2 \hat\lambda }{s^2+s}+2\right)}{2 (\hat\lambda +1)}+O\left(\epsilon ^2\right).
    \end{split}
\end{equation}
For $s$ even, the leading-twist of a spin $s$ operator with $s \geq 1$, is given by: 
\begin{equation}
    \begin{split}
      \tau_{s,\text{min}} & = 2-\frac{\epsilon  \left(2 \hat\lambda +\sqrt{\hat\lambda ^2 \left(s^2+s-2\right)^2-4 \hat\lambda  s \left(s^2+s-10\right) (s+1)+4 s^2 (s+1)^2}+(3 \hat\lambda +2) s (s+1)\right)}{4 (\hat\lambda +1) s (s+1)} \\ & +O\left(\epsilon ^2\right).
    \end{split} \label{epsilon-expansion-higher-spins}
\end{equation}
(We also computed the order $\epsilon^2$ term for finite $s$, but we do not reproduce it here.) Equation \eqref{epsilon-expansion-higher-spins} is plotted in Figure \ref{fig:epsilon-expansion-higher-spins}.

The large $s$ behavior of $\tau_{s,\text{min}}$ for $s$ odd is  
\begin{equation}\begin{split}
    \gamma_s & = -\frac{\hat\lambda \epsilon }{(\hat\lambda +1) s^2} -\frac{\hat\lambda \epsilon^2  \left((4 \gamma -7) \hat\lambda ^2+4 (\gamma -2) \hat\lambda +4 (\hat\lambda +1) \hat\lambda  \log (s)-12\right)}{8 (\hat\lambda +1)^3 s^2} + O(\epsilon^3) \\
    & = -\frac{\hat\lambda \epsilon }{(\hat\lambda +1) s^2} -\frac{ \epsilon^2   \hat\lambda^2  \log  s}{2 (\hat\lambda +1)^2 s^2} + O(\epsilon^3)
    \end{split}
\end{equation}
The smallest twist operator is the stress tensor, which has twist $2-\epsilon$. However, all higher spin operators have twists $2+O(\epsilon)$, and would also contribute to the term of order $\epsilon^2 s^{-2} \log s$ in the above expression.

For operators with even spin, the large-spin behavior of anomalous dimensions depends on whether or not $\hat\lambda$ exceeds $\hat\lambda_*$. For $\hat\lambda<2-\epsilon/3$, the large spin anomalous dimensions, including $\epsilon^2$ terms, are:
\begin{equation}
    \gamma_s \sim -\frac{(\hat\lambda -6) \hat\lambda  \epsilon }{(\hat\lambda -2) (\hat\lambda +1) s^2} + \frac{\hat\lambda  \left(\hat\lambda ^2-6 \hat\lambda -8\right)}{2 (\hat\lambda -2) (\hat\lambda +1)^2 s^2} \epsilon^2 \log s  + O(\epsilon^3).
\end{equation}
and, for $\hat\lambda>2-\epsilon/3$,
\begin{equation}
    \gamma_s \sim -\frac{4 \hat\lambda  \epsilon }{(\hat\lambda -2) (\hat\lambda +1) s^2} + \frac{2 \hat\lambda  (\hat\lambda +2)}{(\hat\lambda -2) (\hat\lambda +1)^2s^2}\epsilon^2 \log s +O(\epsilon^3).
\end{equation}
Above, we only included terms of order $\epsilon^n (\log s)^{n-1}$. These expressions are consistent with the exchange of multiple operators with twists of order $2+O(\epsilon)$ in the four-point function.

The order $\epsilon^2$ correction to $\gamma_s$ for $\hat\lambda=2-\epsilon/3$ is:
\begin{equation}
\begin{split}
    \gamma_s \Big |_{\hat\lambda=\hat\lambda_*} & = -\frac{2}{3s} \epsilon + \\ &  \left(\frac{7 s^4+8 s^3-9 s^2-9 s-2}{9 s^3(s+1) (2 s+1)} - \frac{2 H_{s-1}}{9 s} \right) \epsilon^2 + O(\epsilon^3) \\ 
    & \to -\frac{2}{3s} \epsilon +  \left( \frac{-4 \log (s)-4 \gamma +7}{18 s}\right)\epsilon^2 +O\left(\epsilon^3\right)
\end{split}
\end{equation}
This is consistent with
\begin{equation}
    \gamma_s \sim \left(-\frac{2}{3} \epsilon \right)  s^{-(1-\frac{\epsilon}{3})} = \left(-\frac{2}{3} \epsilon \right)  s^{-\Delta_\sigma},
\end{equation}
as would arise from exchange of the scalar field $\sigma^a$. It would be interesting to understand the large spin behavior of the theory at $\hat\lambda_*$ better via analytic bootstrap methods \cite{Alday:2007mf, Komargodski:2012ek, Fitzpatrick:2012yx,Gopakumar:2016wkt, Dey:2016mcs, Bissi:2022mrs}, assuming they can also be applied to the present disordered CFT, which appears to be unitary.

\bibliography{tensor}
\bibliographystyle{apsrev4-1}

\end{document}